\documentclass[twocolumn]{aastex631}

\shorttitle{Fast Transients}
\shortauthors{Li et al.}
\graphicspath{{./}{figures/}}
\usepackage{textcomp}
\usepackage{comment}
\usepackage{multirow}
\begin{document}
\title{Rapidly Evolving Transients in Archival ZTF Public Alerts}

\author[0000-0002-0096-3523]{Wenxiong Li}
\affiliation{The School of Physics and Astronomy, Tel Aviv University, Tel Aviv 69978, Israel}
\author[0000-0001-7090-4898]{Iair Arcavi}
\affiliation{The School of Physics and Astronomy, Tel Aviv University, Tel Aviv 69978, Israel}
\affiliation{CIFAR Azrieli Global Scholars program, CIFAR, Toronto, Canada}
\author[0000-0002-4534-7089]{Ehud Nakar}
\affiliation{The School of Physics and Astronomy, Tel Aviv University, Tel Aviv 69978, Israel}
\author[0000-0003-3460-0103]{Alexei V. Filippenko}
\affil{Department of Astronomy, University of California, Berkeley, CA 94720-3411, USA}
\author[0000-0001-5955-2502]{Thomas G. Brink}
\affil{Department of Astronomy, University of California, Berkeley, CA 94720-3411, USA}
\author{WeiKang Zheng}
\affil{Department of Astronomy, University of California, Berkeley, CA 94720-3411, USA}
\author[0000-0002-9347-2298]{Marco C. Lam}
\affiliation{The School of Physics and Astronomy, Tel Aviv University, Tel Aviv 69978, Israel}
\author[0000-0002-0675-0887]{Ido Keinan}
\affiliation{The School of Physics and Astronomy, Tel Aviv University, Tel Aviv 69978, Israel}
\author[0000-0003-1325-6235]{Se\'an J. Brennan}
\affiliation{Department of Astronomy, The Oskar Klein Center, Stockholm University, AlbaNova, 106 91 Stockholm, Sweden}
\affiliation{School of Physics, O’Brien Centre for Science North, University College Dublin, Belfield, Dublin 4, Ireland}
\author[0000-0001-9511-6054]{Noi Shitrit}
\affiliation{The School of Physics and Astronomy, Tel Aviv University, Tel Aviv 69978, Israel}

\newcommand\cand{nine}
\newcommand\candtotal{19}
\newcommand\rate{2,400}
\newcommand\auto{\textsc{autophot}}
\newcommand\RETrate{1050$^{+3750}_{-950}$~Gpc$^{-3}$ yr$^{-1}$}

\begin{abstract}
We search the archival Zwicky Transient Facility public survey for rapidly evolving transient (RET) candidates based on well-defined criteria between 2018 May and 2021 December. The search yielded \candtotal\ bona-fide RET candidates, corresponding to a discovery rate of $\sim 5.2$ events per year. Even with a Galactic latitude cut of $20^\circ$, 8 of the 19 events ($\sim 42$\%) are Galactic, including one with a light-curve shape closely resembling that of the GW170817 kilonova (KN). An additional event is a nova in M31. Four out of the 19 events ($\sim 21$\%) are confirmed extragalactic RETs (one confirmed here for the first time) and the origin of 6 additional events cannot be determined. We did not find any extragalactic events resembling the GW170817 KN, from which we obtain an upper limit on the volumetric rate of GW170817-like KNe of $R \le$ \rate~Gpc$^{-3}$~yr$^{-1}$ (95\% confidence).  These results can be used for quantifying contaminants to RET searches in transient alert streams, specifically when searching for kilonovae independently of gravitational-wave and gamma-ray-burst triggers. 

\end{abstract}


\section{Introduction}
\label{sec:intro}
Most extragalactic transients discovered by wide-field optical surveys are thermonuclear and core-collapse (CC) supernovae (SNe) with characteristic timescales (i.e., rest-frame time spent above half of their peak brightness, $t_{1/2}$) longer than 10 days \citep{2020ApJ...904...35P}. Over the last decade, optical surveys 
have discovered a new class of transients with rapidly evolving light curves (on timescales of just a few days). 
Initially, such rapidly evolving transients (RETs) of a variety of spectral types were studied individually, such as SN~2002bj \citep[officially classified as an SN~Ib but with a very peculiar spectrum;][]{2010Sci...327...58P}, SN~2010X \citep[classified as a so-called ``point Ia'' or SN~.Ia;][]{2007ApJ...662L..95B,2010ApJ...723L..98K}, PTF09uj \citep[SN~IIn;][]{2010ApJ...724.1396O}, SN~2005ek \citep[SN~Ic;][]{2013ApJ...774...58D}, iPTF15ul \citep[SN~Ibn;][]{2017ApJ...836..158H}, iPTF16asu \citep[SN~Ic-BL;][]{2017ApJ...851..107W}, and SN~2018kzr \citep{2019ApJ...885L..23M}. The $Kepler$ space telescope, during its K2 mission, also discovered a RET, KSN 2015K, which was observed with a 30~min cadence \citep{2018NatAs...2..307R}. 

With more discoveries, RETs started to be studied in samples. \cite{2014ApJ...794...23D} presented 14 RETs from Pan-STARRS1 \citep[PS1;][]{2016arXiv161205560C}. \cite{2016ApJ...819...35A} reported four especially bright RETs with peak luminosities between those of SNe and superluminous SNe (SLSNe; $M_{\rm peak}\approx -20$~mag) discovered by the Palomar Transient Factory \citep[PTF;][]{2009PASP..121.1395L,2009PASP..121.1334R} and the SN Legacy Survey \citep[SNLS;][]{2006A&A...447...31A}. \cite{2018MNRAS.481..894P} discovered 72 RETs with the Dark Energy Survey \citep[DES;][]{2005IJMPA..20.3121F}. \cite{2020ApJ...904...35P} classified 14 RETs with the Bright Transient Survey of the Zwicky Transient Facility \citep[ZTF;][]{2019PASP..131f8003B,2019PASP..131g8001G,2020ApJ...895...32F}. Recently, \cite{2021arXiv210508811H} presented 42 RETs from Phase I of ZTF. 

Many RETs were found in archival searches and thus have no spectroscopic data. Even those found in real time are challenging to classify spectroscopically given their rapid fading. Those that do have spectra show a variety of characteristics, including blue continua and spectral features associated with normal SNe of various types. This suggests that the RET population is likely heterogeneous in nature.

Several scenarios have been proposed to explain the photometric and spectroscopic properties of RETs, including shock breakout or post-shock cooling emission from a dense wind circumstellar medium \citep[CSM;][]{2010ApJ...724.1396O} or an extended stellar envelope \citep{2013ApJ...774...58D,2017ApJ...851..107W}, the explosion of a stripped massive star \citep{2013ApJ...778L..23T,2014MNRAS.438..318K,2018MNRAS.475.3152K}, white dwarf detonation \citep{2016ApJ...819...35A,2019ApJ...885L..23M}, magnetar spin-down \citep{2016ApJ...819...35A,2017ApJ...851..107W}, and CSM interaction \citep{2016ApJ...819...35A}. However, limited by the low number of events and their heterogeneity, it is challenging to constrain their distribution among different possible physical origins.

The identification of an electromagnetic counterpart to the gravitational wave (GW) event GW170817 \citep[see][and references therein]{2017ApJ...848L..12A} confirmed binary neutron star (BNS) mergers as yet another RET channel, termed a ``macronova'' \citep{2005astro.ph.10256K} or ``kilonova'' \citep[KN; e.g.,][]{1998ApJ...507L..59L,2005ApJ...634.1202R,2010MNRAS.406.2650M,2012ApJ...746...48M,2013ApJ...775...18B,2013ApJ...775..113T,2019LRR....23....1M}. The optical/infrared emission of the electromagnetic transient, known as AT~2017gfo, provided evidence for $r$-process nucleosynthesis in the BNS merger ejecta (e.g., \citealt{2017ApJ...848L..12A, 2017Natur.551...64A,2017ApJ...848L..19C,2017Sci...358.1556C,2017ApJ...848L..17C,2017Sci...358.1570D,2017ApJ...848L..29D,2017Sci...358.1565E,2017SciBu..62.1433H,2017Sci...358.1559K,2017ApJ...850L...1L,2017Natur.551...67P,2017Sci...358.1574S,2017Sci...358.1583K,2017Natur.551...75S,2017ApJ...848L..27T,2017PASJ...69..101U,2017ApJ...848L..24V}; see \citealt{2020PhR...886....1N} for a recent review). However, AT~2017gfo also left many open questions unanswered, such as the nature of its early blue emission \citep[shock cooling, boosted relativistic ejecta, or low-opacity material; see, e.g.,][and references therein]{2018ApJ...855L..23A}, and how common the emission properties of this single event are. Recently, \cite{2022Natur.612..223R} and \cite{2022Natur.612..228T} discovered a KN candidate in the afterglow of an extended-emission gamma-ray burst (GRB; but see \citealt{2022arXiv220610710W} for an alternative explanation). In this case, disentangling the GRB afterglow from the KN emission is not trivial, especially at early times. Therefore, stringent constraints on the early blue emission of this event cannot be readily obtained without making assumptions about the afterglow component.

Transient surveys have been optimizing their search strategies to maximize the probability of a KN discovery without GW or GRB triggers.
\cite{2021MNRAS.500.4213M} presented a marginal KN candidate, AT~2017des, discovered by PS1, though its identification as a KN remains uncertain. 
\cite{2020ApJ...904..155A} searched the first 23 months of ZTF data for candidate KNe, none of which were deemed possible KNe after thorough vetting. 
\cite{2021ApJ...918...63A} introduced a software infrastructure to identify KNe and RETs in real-time ZTF data. They independently found several extragalactic RETs but no KN candidates in 13 months of observations.

Here we perform an independent archival search for unreported KNe and other RETs in Phases I and II of the ZTF public ``Northern Sky Survey'' which covers declinations $\delta \ge -31^{\circ}$ and Galactic latitudes $\mid b_{\rm Gal} \mid > 7^{\circ}$ \citep{2019PASP..131f8003B} from May 2018 to December 2021 (Phase II started in October 2020). Each visible field was scheduled to be observed once in the $g$ band and once in the $r$ band every third night during Phase I and every other night during Phase II. To reduce the number of Galactic variable sources as contaminants, we further restrict the selection here to $\mid b_{\rm Gal} \mid > 20^{\circ}$.

We use the Planck Collaboration et al.\cite{2020A&A...641A...6P} cosmological model throughout. Times are presented in UTC and magnitudes are given in the AB system \citep{1983ApJ...266..713O}.

\section{Methods} \label{sec:method}

We query the ZTF public alert stream between 2018 May 04 and 2021 December 18 using the criteria detailed below. Several public astronomical alert brokers filter, store, and deliver ZTF alerts together with contextual information for each alert. Among them are the Automatic Learning for the Rapid Classification of Events \citep[ALeRCE;][]{2021AJ....161..242F} broker\footnote{\url{https://alerce.online/}}, the Arizona-NOAO Temporal Analysis and Response to Events System
\citep[ANTARES;][]{2018ApJS..236....9N}\footnote{\url{https://antares.noirlab.edu/}}, Lasair \citep{2019RNAAS...3...26S}\footnote{\url{https://lasair-ztf.lsst.ac.uk/}}, FINK \citep{2021MNRAS.501.3272M}\footnote{\url{https://fink-broker.org/}}, and Make Alerts Really Simple (MARS; which has been discontinued)\footnote{\url{https://mars.lco.global/}}. After experimenting with several brokers, we chose ALeRCE for this work (though it could be undertaken with most other brokers as well) owing to its rapid and complete databases, informative website, connections to external archives, and user-friendly \textsc{python} API access. 

\subsection{Selection Criteria} \label{sec:selection}
Since there is no exact definition of RETs, we need to determine a set of selection criteria to efficiently separate them from the stream of roughly 100,000 alerts per day produced by ZTF. These criteria are based on the information provided in the alerts. First, the time span ($t$) during which RETs are detectable is relatively short. We choose the maximum $t$ to be 30 days (shorter than typical SNe, but longer than the KN AT~2017gfo, in the optical bands). Second, we require that the number of positive subtraction detections ({\tt ndet}), with a real-bogus score ({\tt rb}; \citealt{2019PASP..131c8002M}) $> 0.5$, be $> 3$ to exclude likely artifacts with few sporadic detections. 

The most important parameter of a RET is its magnitude rate of change, $\mid \Delta m/\Delta t \mid$, where $\Delta m=m_1-m_2$ and $\Delta t=t_1-t_2$ are the magnitude and time differences between two detections or the first detection and the last nondetection of a candidate. \cite{2020Sci...370.1450D} and \cite{2021NatAs...5...46A} modeled KN light curves from BNS and neutron-star--black-hole (NSBH) mergers, respectively. Their derived 95\% thresholds for the expected decline rates between peak and six days post-peak of NSBH (BNS) KNe are 0.57 (0.60) and 0.39 (0.49) mag~day$^{-1}$ in the $g$ and $r$ bands, respectively. To avoid missing KN-like events, we set the lower threshold of the maximum $\mid \Delta m/\Delta t\mid$ of each candidate to 0.4 mag~day$^{-1}$. We further require $\Delta t > 1$~hr and
$\Delta m> \left(\delta m_1+\delta m_2\right)$, where $\delta m_i$ is the uncertainty in each magnitude measurement. Our approach identifies both rapidly rising and rapidly declining transients, which sets it apart from some previous searches (which focused only on rapid decline rates). 

Finally, since we aim to discover extragalactic transients, we require that there is a galaxy-like object in the vicinity of each candidate. We achieve this via two approaches: (a) cross-matching each candidate position with the Galaxy List for the Advanced Detector Era \citep[GLADE;][]{2018MNRAS.479.2374D} version 2.3 catalog; and (b) cross-matching with galaxy-like sources ({\tt sgscore} $< 0.5$, i.e., more likely a galaxy than a star, based on the star/galaxy separation algorithm of \citealt{2018PASP..130l8001T}) in PS1 images, which cover the entire footprint of the ZTF survey. In both cases, we use a 100\arcsec\ radius for the cross-match.

To summarize, we select events that
\begin{enumerate}
\item are detected for $< 30$ days;
\item have {\tt ndet} $> 3$ detections with a real-bogus score {\tt rb} $> 0.5$;
\item have $\mid \Delta m/\Delta t\mid_{\rm max}\ge0.4 $ mag~day$^{-1}$;
\item 
    \begin{enumerate}
        \item are coincident within 100\arcsec\ with a GLADE galaxy, or
        \item are coincident within 100\arcsec\ with a PS1 object having an {\tt sgscore} $< 0.5$.
    \end{enumerate}
\end{enumerate}

\section{Results} \label{sec:result}
A total of 730 candidates passed the selection criteria detailed in Section \ref{sec:selection}. We examined all of them by eye and found that most have erratic light curves, indicating they are likely subtraction artifacts. We remove these events and are left with 60 candidates with coherent light curves.  
For these candidates, we request forced photometry through the public ZTF forced-photometry service\footnote{https://ztfweb.ipac.caltech.edu/cgi-bin/requestForcedPhotometry.cgi} \citep[$g$ and $r$ bands;][]{2019PASP..131a8003M}, the Asteroid Terrestrial-impact Last Alert System (ATLAS) forced photometry service\footnote{https://fallingstar-data.com/forcedphot/} \citep[$o$ and $c$ bands;][]{2018PASP..130f4505T,2020PASP..132h5002S}, and the All-Sky Automated Survey for SuperNovae (ASAS-SN) forced photometry service\footnote{https://fallingstar-data.com/forcedphot/} \citep[$g$ band;][]{2014ApJ...788...48S,2017PASP..129j4502K} at the median coordinates of each candidate. We request forced photometry from the start of available data for each survey (2018 March for ZTF, 2015 July for ATLAS, and 2013 April for ASAS-SN) to 2021 December. The forced light curves revealed that some candidates are in fact recurrent events (i.e., events with historical outbursts), thus ruling them out as valid extragalactic RETs. 
Furthermore, some candidates rapidly evolve only in certain phases (e.g., early days of an SN), but later evolve more slowly. Thus, we further remove such events by requiring $t_{1/2}\leq15$ days.

\begin{figure*}[!htbp]
\center
\includegraphics[width=1\textwidth]{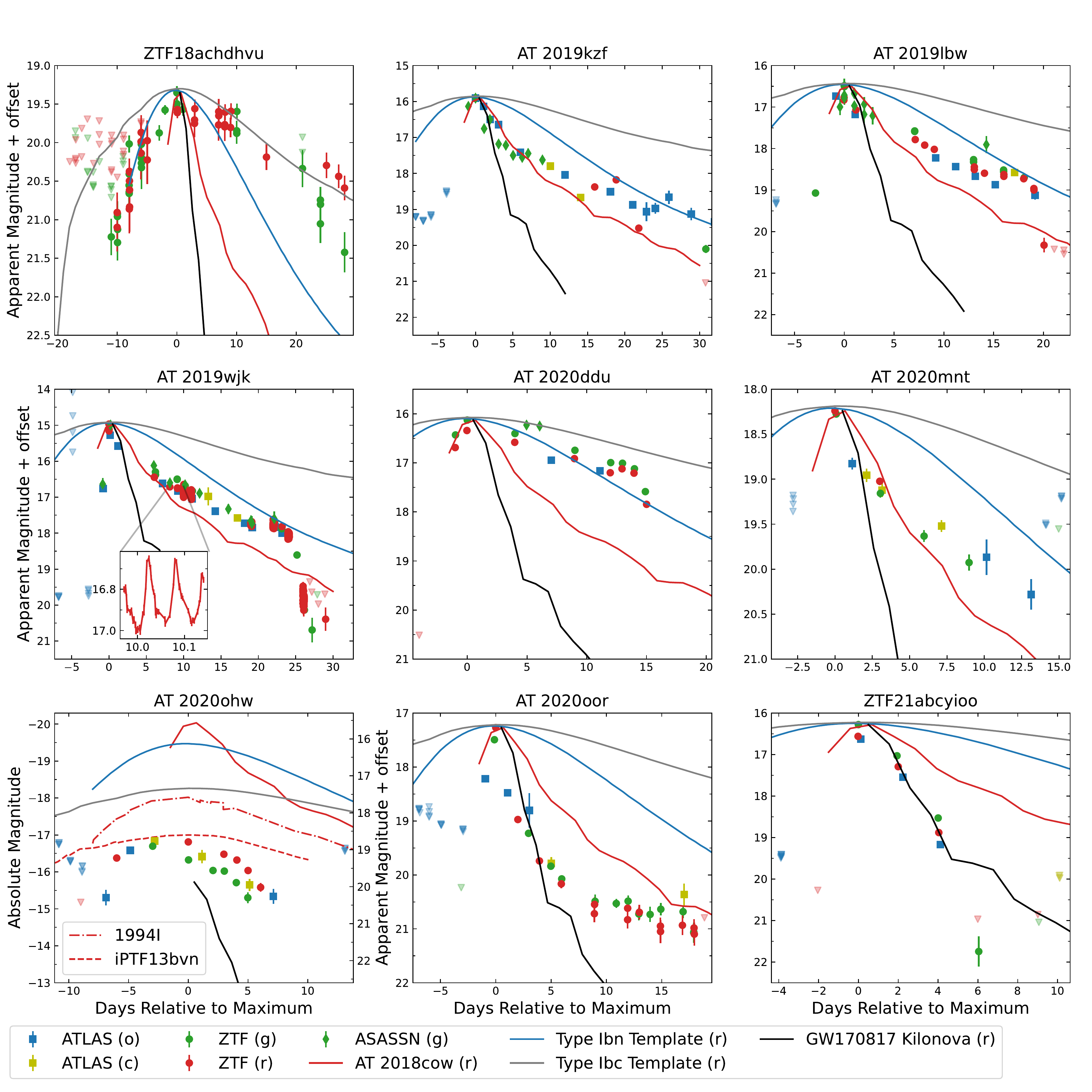}
\caption{Comparison of the light curves of our RET candidates to the SN~Ibc $r$-band template (gray solid line) from \cite{2011ApJ...741...97D}, the SN~Ibn $r$-band template (blue solid line) from \cite{2017ApJ...836..158H}, and $r$-band light curves of a few representative RETs: AT~2018cow \citep[red solid line;][]{2019MNRAS.484.1031P}, and AT~2017gfo \citep[the KN; black solid line;][]{2017PASA...34...69A,2017Natur.551...64A,2017Sci...358.1556C,2017ApJ...848L..17C,2017ApJ...848L..29D,2017Sci...358.1570D,2017ApJ...848L..24V,2017Sci...358.1559K,2017Natur.551...67P,2017Sci...358.1574S,2017Natur.551...75S,2017ApJ...848L..27T,2017PASJ...69..101U}. In all cases we shift the comparison light curves to match the peak apparent magnitude of the RET candidates, except AT~2020ohw, for which we have a spectroscopic redshift and can thus display the light curves in absolute magnitude.  For that event we also plot the $r$-band light curve of Type Ib SN iPTF13bvn \citep[red dashed line;][]{2016A&A...593A..68F} and the $R$-band light curve of Type Ic SN~1994I \citep[red dash-dotted line;][]{1996AJ....111..327R}. In all light curves, triangles denote 3$\sigma$ nondetection limits. A high-cadence oscillating portion of the light curve is seen in the inset of AT~2019wjk. 
}
\label{fig:all-lc}
\end{figure*}

After applying all these filters, a total of \candtotal~candidates remain; their information is listed in Table \ref{tab:pro}. Except for ZTF19accjfgv (SN 2019rta), which appears in the sample of \cite{2021arXiv210508811H}, none of our other candidates are in the samples of \cite{2020ApJ...904..155A}, \cite{2021ApJ...918...63A}, or \cite{2021arXiv210508811H} (see Section \ref{rate} for discussion). We further exclude 10 events with Transient Name Server\footnote{https://www.wis-tns.org/} (TNS) classifications. These include one SN~Ibn (SN~2018fmt), one SN~Icn (SN 2021ckj), one SN~IIb (SN 2019rta, probably detected during its rapidly evolving shock-cooling phase), and seven Galactic objects.  
The light curves of the remaining \cand~unclassified candidates are presented in Figure \ref{fig:all-lc} and are discussed individually below. We calculate peak magnitudes for each candidate using a polynomial fit to their light curves around peak (taking the value of the brighter band if there are observations in more than one band).

\subsection{ZTF18achdhvu}
The rising rate of the light curve of ZTF18achdhvu is similar to the SN~Ibn template from \cite{2017ApJ...836..158H}, while the decline rate resembles that of the SN~Ibc template from \cite{2011ApJ...741...97D}. The $g$-band light curve has a break at $\sim 20$ days after maximum brightness. There is an underlying source, SDSS J122558.34+570930.3, in Sloan Digital Sky Survey \citep[SDSS;][]{2000AJ....120.1579Y} and PS1 
images, 0.4\arcsec\ from the position of ZTF18achdhvu with $g = 21.89 \pm 0.07$~mag, {\tt sgscore} = 0.04 (i.e., a galaxy-like source), and $E(B-V)_{\rm MW} = 0.011$~mag \citep{2011ApJ...737..103S}. It is classified as a galaxy with a photometric redshift of $z_{\rm ph} = 0.215 \pm 0.072$ \citep{2015ApJS..219...12A} by SDSS. 

Considering the high Galactic latitude ($b= 59.6^{\circ}$) and the galaxy-like morphology of the underlying source, it is reasonable to assume ZTF18achdhvu is an extragalactic transient. Taking the photometric redshift, we derive a distance modulus of $40.2^{+0.7}_{-1.0}$~mag, corresponding to Galactic-extinction-corrected absolute peak magnitudes of $M_g \approx -21.0^{+1.0}_{-0.7}$~mag and $M_r \approx -20.8^{+1.0}_{-0.7}$~mag. 

\subsection{AT~2019kzf (ZTF19abdvsjg)}
AT~2019kzf was reported to the TNS through {\it Gaia} alerts 
\citep{2019TNSTR1270....1H} as Gaia19cxh, and was independently detected by ASAS-SN as ASASSN-19px. ATLAS detected the event 15 days before the ZTF discovery. Assuming the first ATLAS detection was at peak brightness, the evolution of the light curve of AT~2019kzf is slightly slower than AT~2018cow while faster than the SN~Ibn template. Its rise is consistent with that of AT~2018cow. A persistent source is identified in SDSS and PS1 images at the position of AT~2019kzf with $g \approx 21.8$~mag and {\tt sgscore} = 0.47. We cannot determine the nature of this source based on current data. If it is not a galaxy but rather a stellar source, AT~2019kzf could be a cataclysmic variable (CV) as proposed by \cite{2022MNRAS.516.2455N}.

    \begin{deluxetable*}{lcDDDcc}
        \tablecaption{RET Candidates that Passed Our Selection Criteria.}
        \label{tab:pro}
        \tablehead{
            \colhead{ZTF Name} & 
            \colhead{Other Name(s)} & 
            \multicolumn2c{$\alpha$ (J2000)} & \multicolumn2c{$\delta$ (J2000)} & \multicolumn2c{$b_{\rm Gal}$} 
            & \colhead{Classification}
            & \colhead{Reference}\\
            \colhead{} & \colhead{} & \multicolumn2c{(degrees)} & \multicolumn2c{(degrees)} & \multicolumn2c{(degrees)} & \colhead{} & \colhead{}
        }
        \decimals
        \startdata
            ZTF18achdhvu & \textemdash & 186.49321 &    57.1585  &    59.6 & Undetermined  & (1) \\
            ZTF19abdvsjg & AT~2019kzf/ & 255.83079 &    12.76519 &    29.7 & Undetermined  & (2) \\
                         & ASASSN-19px &    .      &      .      &      .  &               & \\
            ZTF19abfpwia & AT~2019lbw  & 318.43609 &  $-$6.55881 & $-$34.5 & Undetermined  & (3)  \\
            ZTF19acxmocr & AT~2019wjk  & 109.08012 &    37.45094 &    20.7 & Likely nova  & (4)  \\
            ZTF20aapcjjo & AT~2020ddu  & 290.48945 &    65.91445 &    21.6 & Undetermined  & (5)  \\
            ZTF20abezywi & AT~2020mnt  & 245.72241 & $-$18.37980 &    21.5 & Undetermined  & (6) \\
            ZTF20abkdtvy & AT~2020ohw  & 219.85853 &    13.0189  &    61.1 & Extragalactic & (7) \\
            ZTF20ablatmf & AT~2020oor  & 331.18954 &    12.48679 & $-$33.4 & Undetermined  & (8) \\
            ZTF21abcyioo & \textemdash & 278.26289 &    59.02385 &    25.3 & Galactic      & (9) \\ \hline
            \multicolumn{10}{c}{Classified candidates} \\ \hline
            ZTF18abqbuaj & SN~2018fmt  &  20.25363 & $-$13.86293 & $-$75.0 & SN~Ibn        & (10) \\
            ZTF18absrffm & AT~2018ftw  & 260.23592 &     7.80938 &    23.6 & CV            & (11) \\
            ZTF18abulqzb & AT~2018fzl  & 342.04971 &    16.23608 & $-$37.3 & CV            & (12) \\
            ZTF19accjfgv & SN~2019rta  & 127.20543 &    75.32802 &    32.3 & SN~IIb        & (13) \\
            ZTF19acgfhfd & AT~2019sxc  &  10.75647 &    41.29992 & $-$21.5 & Nova in M31          & (14) \\
            ZTF20acszhad & AT~2020aavs &  76.31652 &  $-$7.28544 & $-$26.9 & Variable Star & (15) \\
            ZTF21aajbgol & SN~2021ckj  & 136.46285 &  $-$8.58543 &    24.8 & SN~Icn        & (16) \\
            ZTF21abexegc & AT~2021plg  & 219.82142 & $-$18.07749 &    37.8 & CV            & (17) \\
            ZTF21abzydnp & \textemdash & 252.51872 &    24.563   &    36.8 & CV            & (18) \\
            ZTF21accbxzk & AT~2021zdk  & 331.92443 &    17.87362 & $-$30.0 & CV            & (19) \\  
        \enddata
        \tablerefs{(1) This work; (2) this work; \cite{2019TNSTR1270....1H}; (3) this work; \cite{2019TNSTR1202....1N}; 
        (4) this work; \cite{2019TNSTR2579....1N}; (5) this work; \cite{2020TNSTR.598....1F}; (6) this work; \cite{2020TNSTR1797....1N}; 
        (7) this work; \cite{2020TNSTR2075....1F}; (8) this work; \cite{2020TNSTR2113....1N}; (9) this work; 
        (10) \cite{2018TNSCR1339....1G,2018TNSCR2146....1F}; (11) \cite{2018TNSCR1404....1F}; (12) \cite{2018TNSCR1397....1W}; 
        (13) \cite{2019TNSCR2407....1D,2021arXiv210508811H}; (14) \cite{2020TNSCR2139....1D}; (15) \cite{2020TNSCR3836....1D}; 
        (16) \cite{2021TNSCR.511....1P,2021TNSAN..71....1P,2021TNSAN..76....1G}; (17) \cite{2021TNSCR2033....1D}; (18) \cite{2016MNRAS.456.4441C}; (19) \cite{2021TNSCR3242....1P}.}
    \end{deluxetable*}
\subsection{AT~2019lbw (ZTF19abfpwia)}\label{sec:2019lbw}
AT~2019lbw was reported to the TNS by \cite{2019TNSTR1202....1N}. It has a very rapid rise with $\mid \Delta m/\Delta t\mid= 0.9$~mag~day$^{-1}$ and a post-peak decline rate between that of AT~2018cow and the SN~Ibn template. Approximately 20 days post-peak, the $r$-band light curve shows a steep drop. There is no obvious host-galaxy candidate or another underlying optical source in the vicinity of AT~2019lbw. We obtained deep imaging at the position of AT~2019lbw using Keck~I/LRIS \citep{1995PASP..107..375O,2010SPIE.7735E..0RR} on 2022 September 21.

Forced point-spread-function (PSF) photometry at the position of AT~2019lbw was performed using the \auto~pipeline \cite{2022A&A...667A..62B}\footnote{https://github.com/Astro-Sean/autophot}, with a PSF model built from bright, isolated sources in the image. Optical magnitudes were calibrated against SDSS field stars. 
We detect a source with $g=24.33\pm0.19$~mag and $r=24.25\pm0.34$~mag (Figure \ref{fig:AT2019lbw_keck}) at the position of AT~2019lbw, but cannot determine if it is a Galactic or extragalactic object. 

The Catalina Real-time Transient Survey \citep[CRTS;][]{2009ApJ...696..870D} detected an outburst which was 1\arcsec\ away from AT~2019lbw on 2009 May 27 with an unfiltered magnitude of $\sim 17.5$, $\sim 1$~mag fainter than the peak of AT~2019lbw. This event might be related to AT~2019lbw, with both events being flares of the same Galactic source whose proper motion between 2009 to 2019 is $\sim1$\arcsec. Alternatively, both events could be extragalactic transients from an undetected host galaxy, or they could be two spatially coincident events without any physical relation. Since we do not have proper-motion measurements or spectroscopic observations of the source, we cannot distinguish between these scenarios. 

\subsection{AT~2019wjk (ZTF19acxmocr)}
\begin{figure*}
\center
\includegraphics[width=1\textwidth]{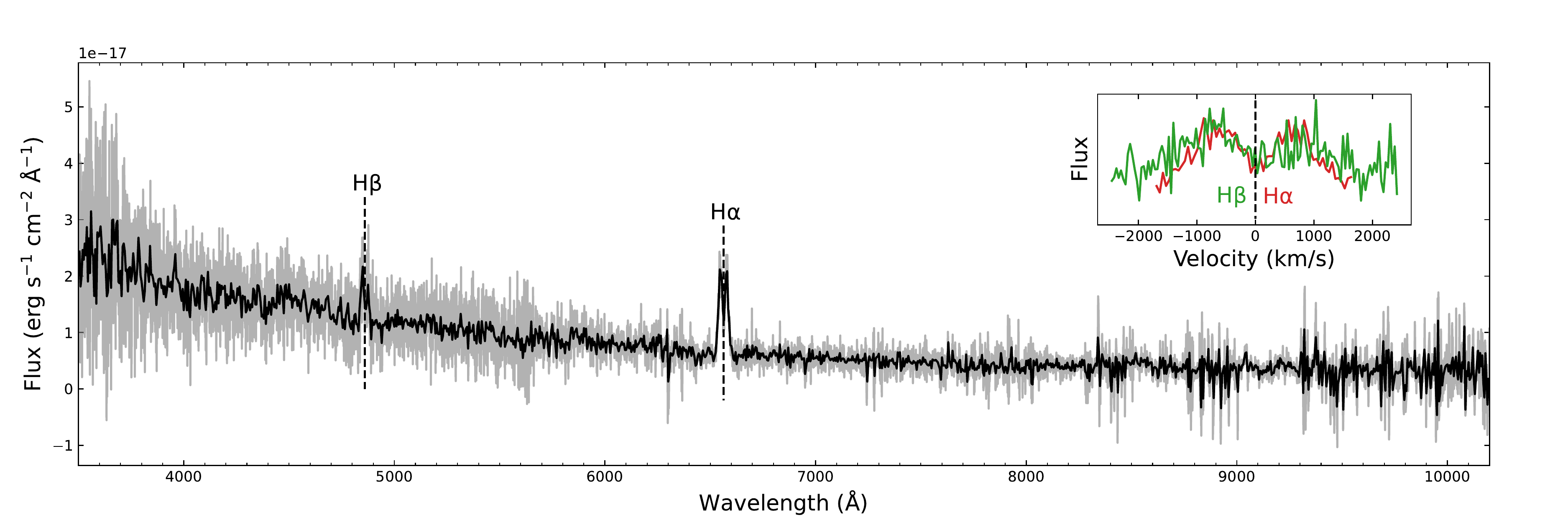}
\caption{The observed (light gray) and binned (black) spectrum taken at the position of AT~2019wjk in 2022. The double-peaked $z \approx 0$ H$\alpha$ and H$\beta$ emission lines (black dashed lines) are consistent with an accreting Galactic source, making the 2019 outburst a possible nova. The continuum-normalized spectra of H$\alpha$ and H$\beta$ in velocity space are displayed in the inset.}
\label{fig:2019wjk_s}
\end{figure*}

AT~2019wjk was reported to the TNS by \cite{2019TNSTR2579....1N}.
It met our RET criteria owing to its $g$-band drop of 1.51~mag~day$^{-1}$, at $\sim 25$ days after peak light. ATLAS $o$-band and ASAS-SN $g$-band data reveal a very rapid rise of $\sim 1.5$~mag~day$^{-1}$, faster than that of AT~2018cow. With a low Galactic latitude of $b_{\rm Gal} = 20.7^\circ$ (very close to our cutoff), AT~2019wjk is located within the ZTF high-cadence Galactic-plane survey fields \citep{2021MNRAS.505.1254K}, with the data during the outburst published in ZTF Public Data Release 7\footnote{https://www.ztf.caltech.edu/ztf-public-releases.html} \citep{2019PASP..131a8003M}. The $\sim 5$~min cadence $r$-band light curve from this survey, taken between 10 and 11 days after peak, clearly shows an oscillation pattern between 16.6 and 17.0~mag (see inset in Figure~\ref{fig:all-lc}), typical of post-peak light curves of novae \citep{2010AJ....140...34S,2012JAVSO..40..582M}.
A persistent source is identified in PS1 images at the position of AT~2019wjk with $g = 21.74 \pm 0.04$~mag and {\tt sgscore} = 0.30. To determine the nature of this source, we obtained a Keck~I/LRIS spectrum of the position of AT~2019wjk on 2022 March 03. The spectrum was processed with the \textsc{LPipe} data-reduction pipeline \citep{2019PASP..131h4503P}. 
Figure \ref{fig:2019wjk_s} displays the reduced spectrum and the binned spectrum. It shows double-peaked Balmer lines around a redshift of $z=0$. The low Galactic latitude, rapid oscillations, and redshift zero double spectral lines are all consistent with AT~2019wjk being a nova outburst of a CV.

\subsection{AT~2020ddu (ZTF20aapcjjo)} \label{2022ddu}
AT~2020ddu was reported to the TNS by \cite{2020TNSTR.598....1F}. It displays a rapid rise, similar to that of AT~2018cow. After peak light, however, it evolves slower, following the SN~Ibn template until $\sim 15$ days after peak when the light curve shows a drop in both the $g$ and $r$ bands. The decline rate during this drop is 0.59~mag~day$^{-1}$ in $r$. No source is detected in PS1 at the position of AT~2020ddu. We obtained deep imaging at this position using Keck~I/LRIS on 2022 July 01. The data-reduction process is the same as that described in Section \ref{sec:2019lbw}.

We marginally detect a source with $g=24.1\pm0.3$~mag and $r=24.3\pm0.5$~mag (Figure \ref{fig:AT2020ddu_keck}) but cannot determine whether it is a Galactic or extragalactic object.
\subsection{AT~2020mnt (ZTF20abezywi)}
AT~2020mnt was reported to the TNS by \cite{2020TNSTR1797....1N}. Assuming AT~2020mnt was discovered at peak brightness, the decline rate during the first three days from discovery is comparable to that of AT~2018cow. After five days the decline becomes slower and more similar to that of the SN~Ibn template. There is an underlying source visible in archival PS1 images 0.39\arcsec\ from AT~2020mnt that has $g\approx 24$~mag, $r\approx 23$~mag \citep{2013A&A...554A.101B} and {\tt sgscore} = 0.23. We cannot determine the nature of this source based on current data. 

\subsection{AT~2020ohw (ZTF20abkdtvy)}
AT~2020ohw was reported to the TNS by \cite{2020TNSTR2075....1F} and is the only event in our sample with an archival spectrum of its host galaxy.  
The host, SDSS~J143926.04+130108.2, is at $z = 0.02886 \pm 0.00001$ \citep{2010A&A...514A.102T}, with AT~2020ohw located at its center. The corresponding luminosity distance is $134 \pm 9$~Mpc and the distance modulus is $35.6 \pm 0.2$~mag. The peak absolute $r$-band magnitude of AT~2020ohw after Galactic extinction correction 
is $M_r=-16.9 \pm 0.3$~mag. The light-curve decline rate of AT~2020ohw is between those of AT~2018cow and the SN~Ibn template, with a maximum $g$-band decline rate of $\mid \Delta m/\Delta t\mid_{\rm max} = 0.50$~mag~day$^{-1}$ at $\sim 5$ days after peak. Its luminosity, however, is much lower than those of AT~2018cow and the SN~Ibn template. In Figure \ref{fig:all-lc}, we further compare the $r$-band light curve of AT~2020ohw with that of iPTF13bvn \citep{2016A&A...593A..68F}, a low-luminosity SN~Ib, and with the $R$-band light curve of SN~1994I \citep{1996AJ....111..327R}, a rapidly evolving SN~Ic. AT~2020ohw is fainter than SN~1994I, although both events have similar rising rates. The peak of the light curve of AT~2020ohw is similar to that of iPTF13bvn but AT~2020ohw rises faster to peak. We propose that AT~2020ohw is a RET, possibly a relatively low-luminosity and rapidly evolving SN~Ib/c.

\subsection{AT~2020oor (ZTF20ablatmf)}\label{sec:2020oor}
AT~2020oor was reported to the TNS by \cite{2020TNSTR2113....1N}. 
The rise has a similar rate to that of AT~2018cow, and the initial decline is even faster. In the first four days after peak, the decline rate is similar to that of AT~2017gfo, but later slows down and even flattens. No source is detected in the vicinity of AT~2020oor in archival images. We obtained $g$ and $r$ imaging at the position of AT~2020oor 
with Keck~I/LRIS on 2022 June 28. The data-reduction process is the same as that described in Section \ref{sec:2019lbw}. A possible source is detected by \auto~ with a signal-to-noise ratio of $\sim 2$ in the $g$ and $r$ images (Figure \ref{fig:AT2020oor_keck}). We perform artificial source injection using the PSF model of each image to determine the limiting magnitudes of the image (see \cite{2022A&A...667A..62B} for further details), obtaining a 3$\sigma$ upper limit of 25.00~mag in $g$ and 24.95~mag in $r$. Unfortunately, we cannot determine the nature of this source based on the current data. 

\subsection{ZTF21abcyioo}
ZTF21abcyioo is the most promising KN candidate among our sample, having a light curve similar in shape to that of AT~2017gfo (Figure \ref{fig:all-lc}). The last ZTF $g$-band detection and adjacent upper limit, at six days after discovery, show an even faster late-time decline than that of AT~2017gfo. The Galactic-extinction-corrected $g-r$ color of ZTF21abcyioo is between $\sim -0.2$ and $-0.3$~mag, not as red as that of AT~2017gfo which starts from $\sim 0$ mag and evolves redward to $\sim 1.5$~mag at five days after peak. There is an underlying source in archival PS1 images at the position of ZTF21abcyioo that has $g = 21.71 \pm 0.04$~mag, $r = 20.87 \pm 0.03$~mag \citep{2016arXiv161205560C}, and {\tt sgscore} = 0.38. The Galactic latitude of ZTF21abcyioo is $b_{\rm Gal} = 25.3^\circ$, close to the limits of our search. The {\it Gaia} Data Release 3 (DR3) does not provide a measurable proper motion of this source \citep{2022arXiv220800211G}.  
The DESI Legacy Imaging Surveys catalog classifies the archival source as a stellar object but does not provide an ellipticity measurement \citep{2019ApJS..245....4Z,2019AJ....157..168D}. 

\begin{figure*}
\center
\includegraphics[width=1\textwidth]{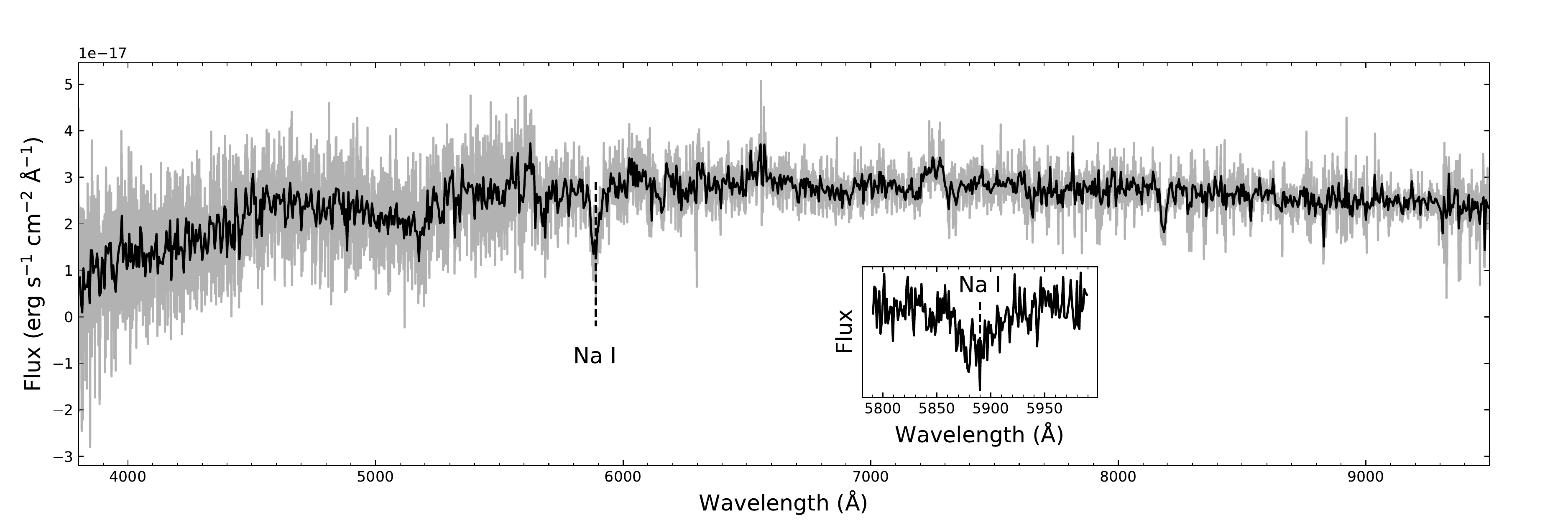}
\caption{The observed (light gray) and binned (black) spectrum taken at the position of ZTF21abcyioo in 2022. The $z = 0$ Na~{\sc i}~D absorption doublet is labeled (dashed line), and the continuum-normalized spectrum of it is displayed in the inset.}
\label{fig:ZTF21abcyioo_s}
\end{figure*}

To reveal the nature of ZTF21abcyioo, we obtained spectra at the position of ZTF21abcyioo with Keck~I/LRIS on 2022 September 23. The data were optimally extracted with \textsc{aspired}\footnote{https://github.com/cylammarco/ASPIRED} \citep{2021arXiv211102127L, aspired_zenodo}. We identify the prominent absorption doublet of sodium Na~{\sc i}~D $\lambda\lambda$5890 (D$_2$), 5896 (D$_1$) (Figure \ref{fig:ZTF21abcyioo_s}) at $z = 0$. According to the relation from \cite{2012MNRAS.426.1465P}, the Galactic extinction of $E(B-V) = 0.04$~mag at this position \citep{2011ApJ...737..103S} corresponds to an equivalent width of EW(D$_1$ + D$_2$) $= 0.39~\pm~0.07$~\AA\ for the Na~{\sc i}~D doublet. 
However, we measure the observed EW(D$_1$ + D$_2$) from the spectrum to be $\sim 17.6$~\AA, much larger than the Galactic contribution. Therefore, we suggest that the major contribution to these features is from the underlying source, making it a $z=0$ Galactic object. Unfortunately, we cannot determine the nature of the source or its outburst. To calculate the probability that the spatial coincidence between ZTF21abcyioo and a foreground Galactic source could be due to chance, we count the number of stellar sources in the PS1 catalog down to its limiting magnitude of $g = 23.3$~mag \citep{2016arXiv161205560C} in a $0.1^\circ$ radius region around ZTF21abcyioo, finding 51 sources. Conservatively, assuming we would associate ZTF21abcyioo with a stellar source if it were within a radius of 2\arcsec\ from it, we find that the probability of a chance coincidence in this field is only 2\%. This reaffirms our conclusion that ZTF21abcyioo is probably of Galactic origin.

\section{Discussion}\label{rate}
Here we discuss the implications of our search results for future surveys. First, we wish to quantify potential contaminants to RET searches in general, and to KN searches in particular (both following, and independently of, GW alerts). During the 3.75~yr period of our search, we found 60 RET candidates with coherent light curves (see Section \ref{sec:result}), or 16 events per year. Since we required a limit on the duration of RET candidates, transients that start rising quickly but are detected for a total duration of more than 30 days in the ZTF survey are not included in our sample. Those transients could be contaminants to real-time searches. Therefore, our estimate of 16 events per year can only be considered as a lower limit for what can be found in real-time searches. Of those, we identified 19 valid RET candidates in 3.6~yr, corresponding to 5.2 events per year.

\begin{figure}
\center
\includegraphics[width=1\columnwidth]{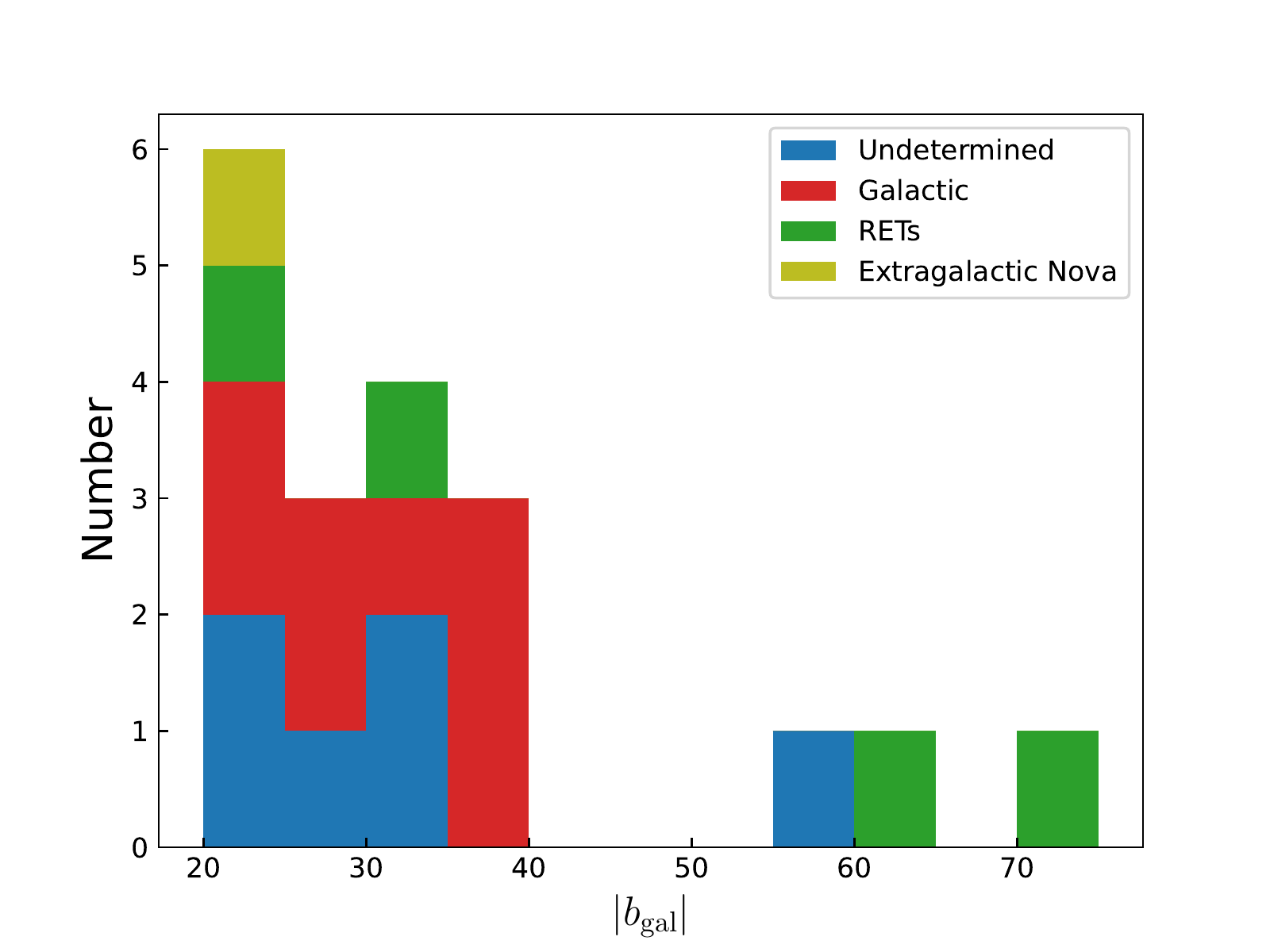}
\caption{The distribution of $\vert b_{\rm gal} \vert$ for undetermined events (blue), Galactic objects (red), RETs (green), and an extragalactic nova (yellow). Significant Galactic contamination extends out to $\vert b_{\rm gal} \vert$ = 40$^\circ$.}
\label{fig:b}
\end{figure}

The distributions of Galactic latitude, $\vert b_{\rm Gal} \vert$, for our events are displayed in Figure \ref{fig:b}.
As expected, Galactic objects are concentrated in lower $b_{\rm Gal}$; however, there is still a significant number of these events out to $\vert b_{\rm Gal} \vert = 40^\circ$. Most undetermined events are also at relatively low $\vert b_{\rm Gal} \vert$, indicating that they are likely Galactic as well, with only one above $40^\circ$. RETs are more evenly distributed compared to the other two groups. These results can help fine tune future search criteria and contamination expectations.

Next, we compare our search results with previous studies. Only one of our \candtotal~RET candidates, SN~2019rta, overlaps with samples found by previous searches \citep{2021arXiv210508811H}. One main reason that we do not recover some of the existing candidates is that we use data only from the public ZTF survey, which is a subset of the full survey data available to \cite{2020ApJ...904..155A}, \cite{2021ApJ...918...63A}, and \cite{2021arXiv210508811H}. Additionally, there are several differences between our search criteria and those in the literature. \cite{2020ApJ...904..155A} and \cite{2021ApJ...918...63A} selected only rapidly declining transients while our search selected both rapidly rising and rapidly declining events. Also, \cite{2020ApJ...904..155A} and \cite{2021ApJ...918...63A} rejected candidates with a total time span of $> 14$~days, which is much shorter than our threshold of 30 days. \cite{2021arXiv210508811H} required that the light curves of the candidates be well sampled; there had to be an observation within 5.5 days both before and after peak brightness in both the $g$ and $r$ bands,  a criterion that some of our candidates do not fulfill. Finally, all searches in the literature ended before 2020 November; thus, all of our candidates from after that epoch could not have been identified previously.

Our search yielded six undetermined events, four RETs (of which three were previously known, and AT~2020ohw is newly identified here), but no valid KN candidates. We can utilize these results to constrain the volumetric rate of RETs and KNe. We use {\tt simsurvey} \citep{2019JCAP...10..005F}, a software package for simulating transient light curves detected based on a given survey strategy, to calculate the occurrence rates. The details of the ZTF CCDs and filters are embedded in {\tt simsurvey}. We use the actual observation schedule of the ZTF public survey from our search period using the ZTF Public Data Release 14 \citep{2019PASP..131a8003M} which includes coordinates, time, and filter information for each exposure. 
Milky Way extinction was applied to each survey pointing based on the measurements from \cite{1998ApJ...500..525S}. 

For KNe, we use the best-fit radiative transport parameters 
derived by \textsc{possis} \citep{2019MNRAS.489.5037B} for AT~2017gfo in \cite{2020Sci...370.1450D} and assume the distribution of viewing angle $i$ is uniform in cos($i$), following \cite{2020ApJ...904..155A}. We injected the simulated spectra into {\tt simsurvey} in a redshift range $z~\in$ (0, 0.03) with a given rate. We filter the simulated light curves with the criteria described in Section \ref{sec:selection}.  
According to Poisson small-number statistics, using the method of \cite{1986ApJ...303..336G}, and given that we detected no KNe in our search, we find that the 95\% confidence upper limit to the number of actual KNe during our search period is three. Therefore, our upper limit on the KN rate is the rate at which three KNe are recovered from the simulation.
Combining the survey schedule with the KN model, we derive a rate of $R~\le$ \rate~Gpc$^{-3}$~yr$^{-1}$. This is consistent with but less constraining than the rate limit of \cite{2021ApJ...918...63A} ($R~\le$ 900~Gpc$^{-3}$~yr$^{-1}$ at 95\% confidence) for GW170817-like KNe.

The rate calculation for RETs is slightly more complicated. Our search yielded four RETs of varying peak $r$-band magnitudes ($-18.8$ for the Type~Ibn SN 2018fmt, $-17.4$ for the Type IIb SN 2019rta, $-$19.8 for the Type Icn SN 2021ckj, and $-$16.6 for AT 2020ohw, the new RET identified here). Since the assumed light-curve luminosity has a strong effect on the rate estimates, and since we do not know the true luminosity function of RETs, here we estimate the rate only of AT~2020ohw-like RETs (i.e., with a similar peak luminosity). {\tt simsurvey} requires a time-dependent spectral energy distribution (SED) template, which we do not have for AT~2020ohw; we therefore use AT~2018cow as a model, building the SED template from the AT~2018cow spectra of \cite{2019MNRAS.484.1031P} and \cite{2021ApJ...910...42X} assuming the emission is isotropic. We then scale the templates to match the light curve of AT~2020ohw around peak and use that as input for {\tt simsurvey}. The process described above is followed to calculate the volumetric rate of RETs based on 
a single event discovered in our search. With a 95\% confidence interval, the number of actual AT~2020ohw-like RETs during our search period is $1^{+3.74}_{-0.95}$, and the resulting volumetric rate estimate is \RETrate. This is a lower limit, even just on the rate of AT~2020ohw-like RETs, since we do not know if any of the six undetermined events in our search are also of this type. This result is therefore broadly consistent with the $\ge 1000$~Gpc$^{-3}$~yr$^{-1}$ rate calculated by \cite{2018MNRAS.481..894P} for FBOTs peaking between $M_g = -15.8$ and $M_g = -22.2$~mag, and the 4800--8000~Gpc$^{-3}$~yr$^{-1}$ rate calculated by \cite{2011ApJ...741...97D} for FBOTs peaking between $M_g = -16.5$ and $M_g = -20$~mag.

\section{Conclusions}\label{conclusion}

Searching the archival ZTF public survey for RET candidates based on well-defined criteria between 2018 May and 2021 December yielded \candtotal~bona-fide RET candidates, corresponding to a discovery rate of five events per year. To try to determine their physical origin, we obtained observations of some candidates using Keck~I/LRIS. Of \cand~previously unclassified RET candidates, we find that one is extragalactic, two are Galactic, and the nature of the remaining six events remains undetermined. We derive a volumetric rate of RETs similar to our one extragalactic confirmed candidate (which peaks at $M_r=-16.6$~mag) of $R$ = \RETrate. No valid KN candidates were detected, from which we are able to obtain an upper limit on the volumetric rate of GW170817-like KNe of $R~\le$ \rate~Gpc$^{-3}$~yr$^{-1}$. 

This work quantifies the contamination to KN and RET searches. Even with a Galactic latitude cut of $20^\circ$, at least 8 and possibly up to 14 of the \candtotal~events ($\sim 42$--74\%) are Galactic, with a substantial number of them extending out to a Galactic latitude of $40^\circ$. This includes one Galactic event with a light-curve shape closely mimicking that of the GW170817 KN. Between 4 and 10 of the \candtotal~events ($\sim 21$--53\%) are RETs. These results can be used for estimating contamination rates in searches for RETs in transient alert streams and specifically for searching for KNe independently of GW and GRB triggers.

\section*{Acknowledgments}
W.L. is supported by an Israel Science Foundation (ISF) grant (number 2752/19) and a European Research Council (ERC) grant (JetNS) under the European Union's Horizon 2020 research and innovation program.
I.A. is a CIFAR Azrieli Global Scholar in the Gravity and the Extreme Universe Program and acknowledges support from that program, from the ERC (grant agreement number 852097), from the ISF (grant number 2752/19), from the United States -- Israel Binational Science Foundation (BSF), and from the Israeli Council for Higher Education Alon Fellowship.
M.C.L. is supported by the ERC (grant agreement number 852097).
S.J.B. acknowledges support from Science Foundation Ireland and the Royal Society (RS-EA/3471).
A.V.F.'s group at UC Berkeley has been supported by the Christopher R. Redlich Fund, Alan Eustace (W.Z. is a Eustace Specialist in Astronomy), Frank and Kathleen Wood (T.G.B. is a Wood Specialist in Astronomy), and numerous individual donors.

This work is based on observations obtained with the Samuel Oschin 48-inch Telescope at Palomar Observatory as part of the Zwicky
Transient Facility project. ZTF is supported by the U.S. National Science Foundation (NSF) under grant AST-1440341 and a
collaboration including Caltech, IPAC, the Weizmann Institute for Science, the Oskar Klein Center at Stockholm University, the
University of Maryland, the University of Washington, Deutsches Elektronen-Synchrotron and Humboldt University, Los Alamos
National Laboratories, the TANGO Consortium of Taiwan, the University of Wisconsin at Milwaukee, and Lawrence Berkeley
National Laboratories. Operations are conducted by COO, IPAC, and UW.

We thank Eric C. Bellm for sharing the ZTF observed pointing history used in this paper.

Some of the data presented herein were obtained at the W. M. Keck Observatory, which is operated as a scientific partnership among the California Institute of Technology, the University of California, and the National Aeronautics and Space Administration (NASA). The Observatory was made possible by the generous financial support of the W. M. Keck Foundation. The authors wish to recognize and acknowledge the very significant cultural role and reverence that the summit of Maunakea has always had within the indigenous Hawaiian community.  We are most fortunate to have the opportunity to conduct observations from this mountain.

The Pan-STARRS1 Surveys (PS1) and the PS1 public science archive have been made possible through contributions by the Institute for Astronomy, the University of Hawaii, the Pan-STARRS Project Office, the Max-Planck Society and its participating institutes, the Max Planck Institute for Astronomy, Heidelberg and the Max Planck Institute for Extraterrestrial Physics, Garching, The Johns Hopkins University, Durham University, the University of Edinburgh, the Queen's University Belfast, the Harvard-Smithsonian Center for Astrophysics, the Las Cumbres Observatory Global Telescope Network Incorporated, the National Central University of Taiwan, the Space Telescope Science Institute, NASA under grant NNX08AR22G issued through the Planetary Science Division of the NASA Science Mission Directorate, NSF grant AST-1238877, the University of Maryland, Eotvos Lorand University (ELTE), the Los Alamos National Laboratory, and the Gordon and Betty Moore Foundation.

This research has made use of the VizieR catalogue access tool, CDS, Strasbourg, France (DOI: 10.26093/cds/vizier). The original description of the VizieR service was published in 2000, A\&AS 143, 23.

\facilities{Keck~I/LRIS \citep{1995PASP..107..375O,2010SPIE.7735E..0RR}}

\software{\textsc{Astropy} \citep{2013A&A...558A..33A}, \textsc{Aspired} \citep{2021arXiv211102127L, aspired_zenodo}, \auto~ \cite{2022A&A...667A..62B}, \textsc{LPipe} \citep{2019PASP..131h4503P}, \textsc{Matplotlib} \citep{2021zndo....592536C}, \textsc{Pandas} \citep{2022zndo...3509134R}, \textsc{simsurvey} \citep{2019JCAP...10..005F}.}

\appendix
\section{Deep Photometry}
\begin{figure*}
\center
\includegraphics[width=1\textwidth]{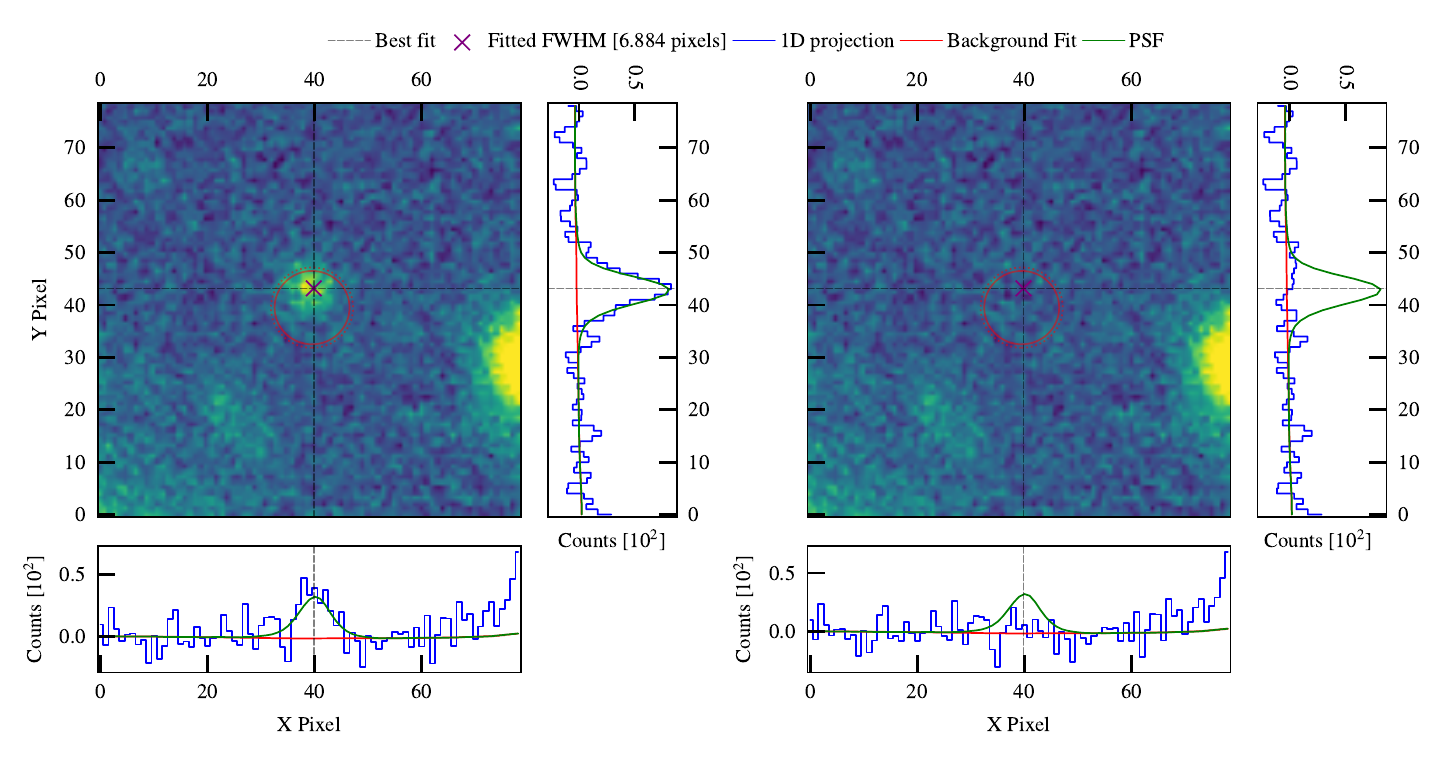}
\includegraphics[width=1\textwidth]{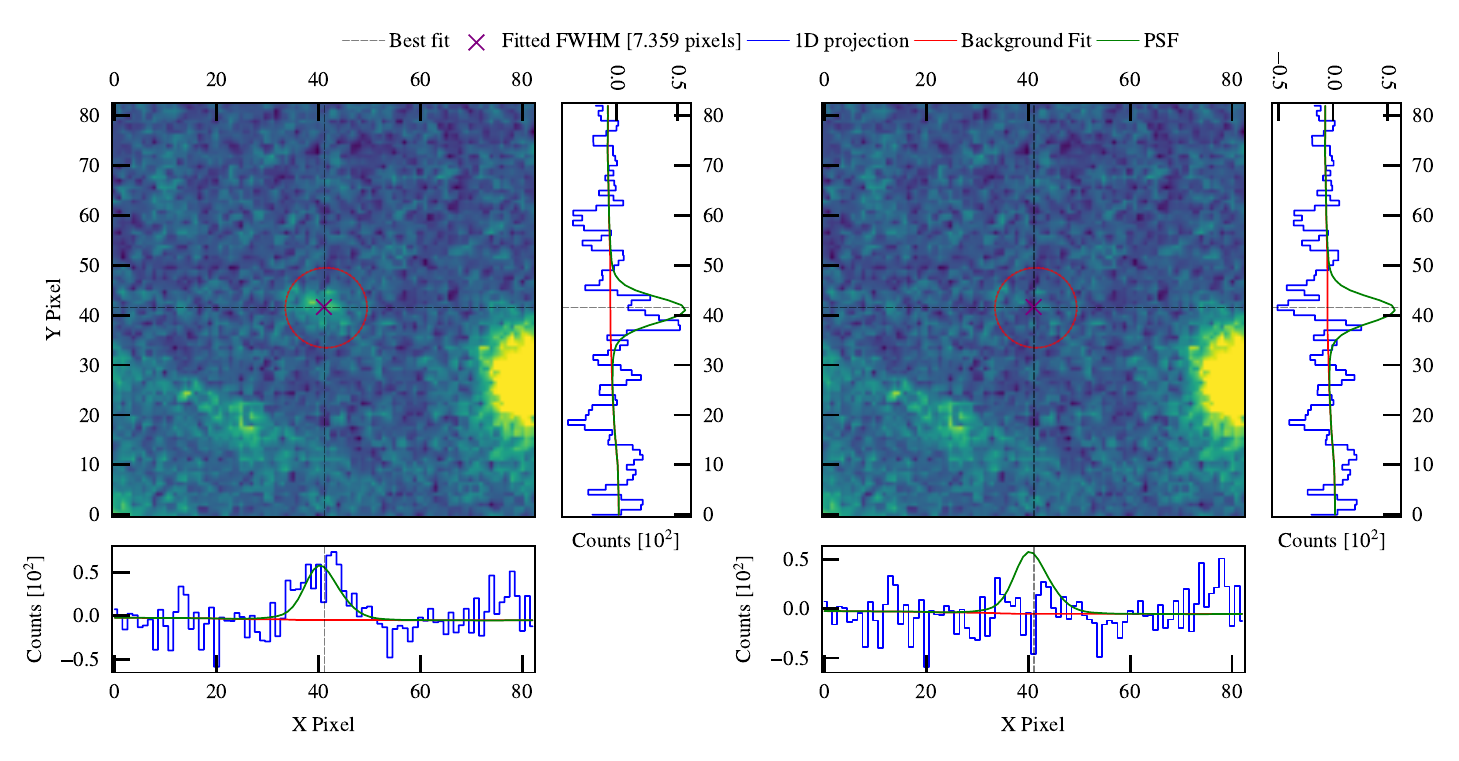}
\caption{PSF subtraction of the source at the position of AT~2019lbw in $g$ (upper panels) and $r$ (lower panels) imaging, produced with \auto. The main panels show a cutout of the target position before (left) and after (right) PSF subtraction, while projections along the $X$ and $Y$ axes are also shown for each panel. }
\label{fig:AT2019lbw_keck}
\end{figure*}

\begin{figure*}
\center
\includegraphics[width=1\textwidth]{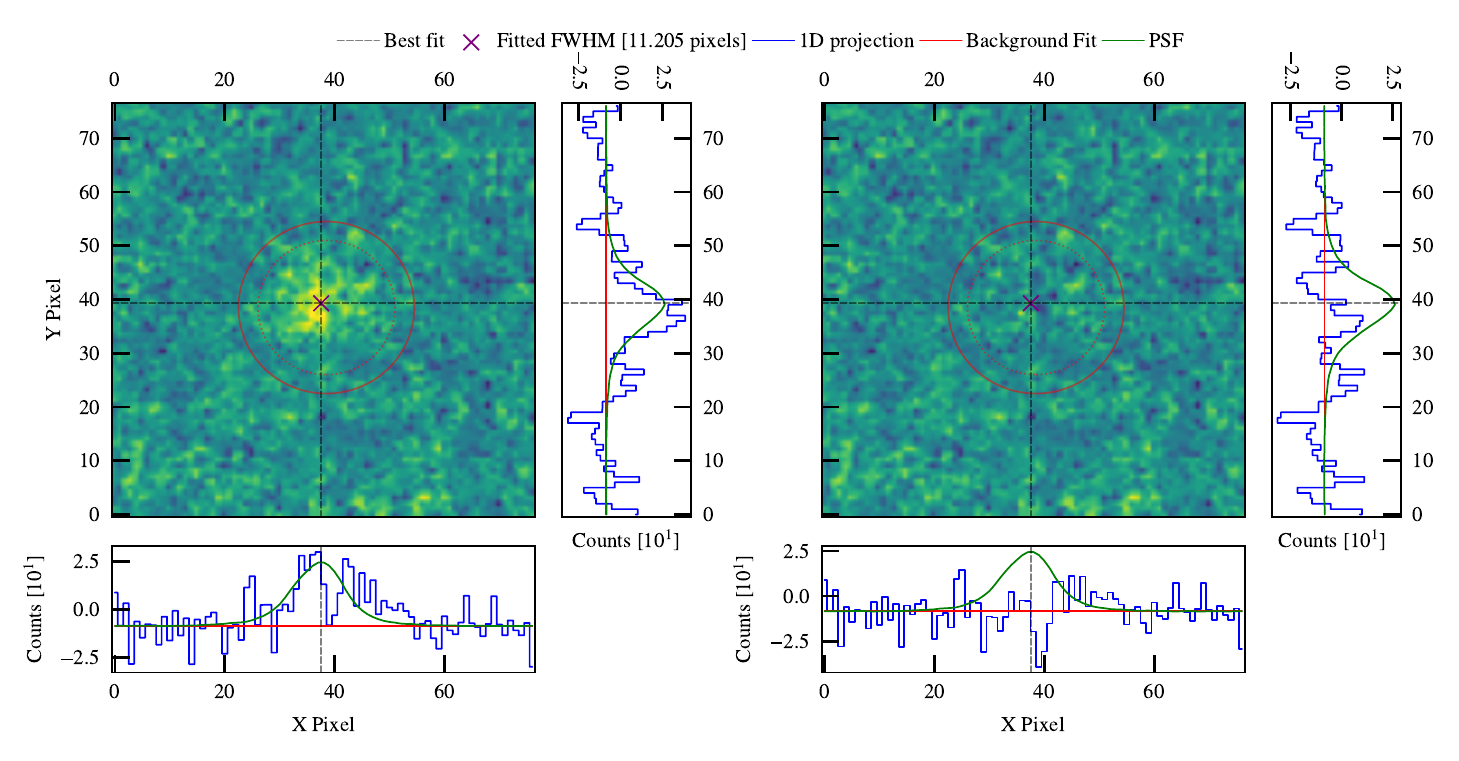}
\includegraphics[width=1\textwidth]{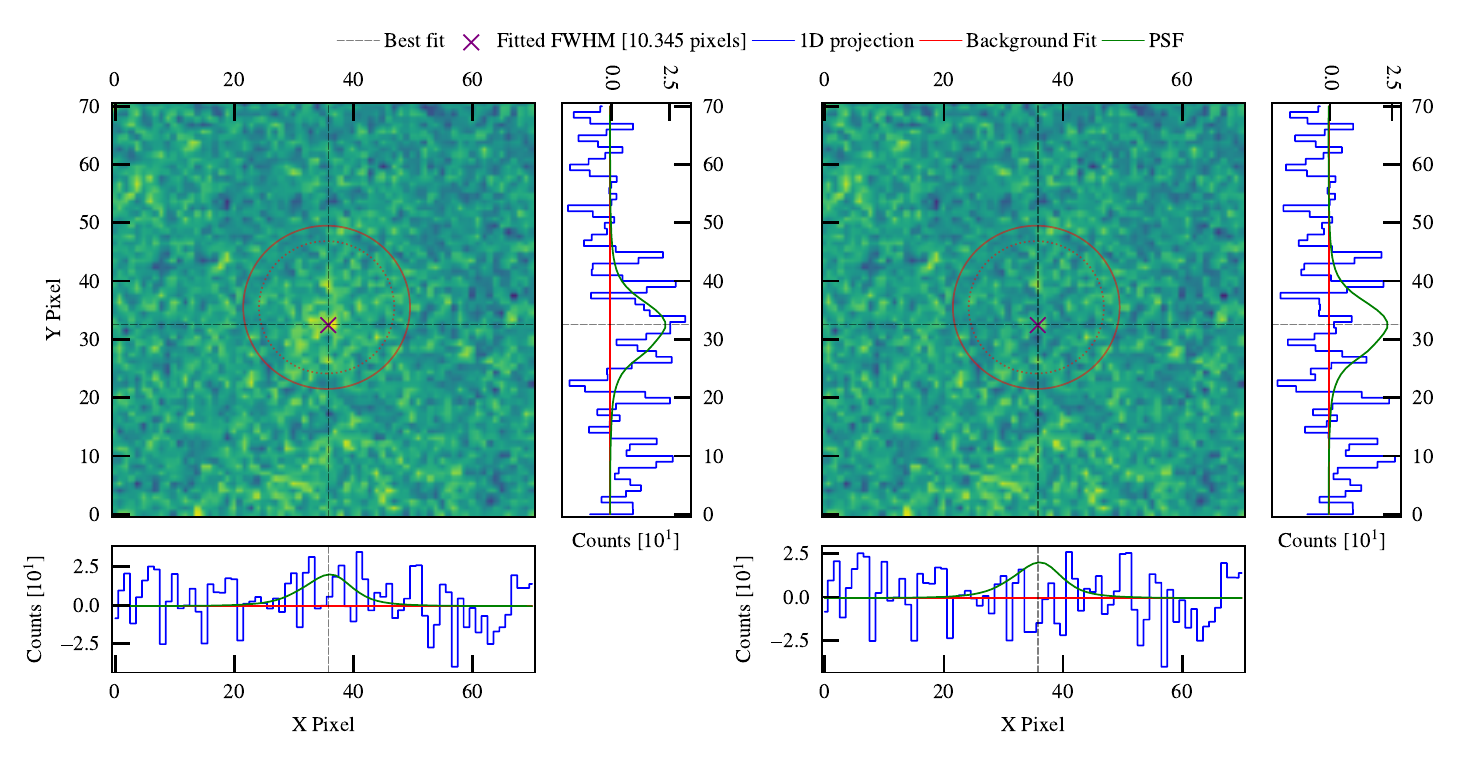}
\caption{Same as Figure \ref{fig:AT2019lbw_keck} but for AT~2020ddu.}
\label{fig:AT2020ddu_keck}
\end{figure*}

\begin{figure*}
\center
\includegraphics[width=1\textwidth]{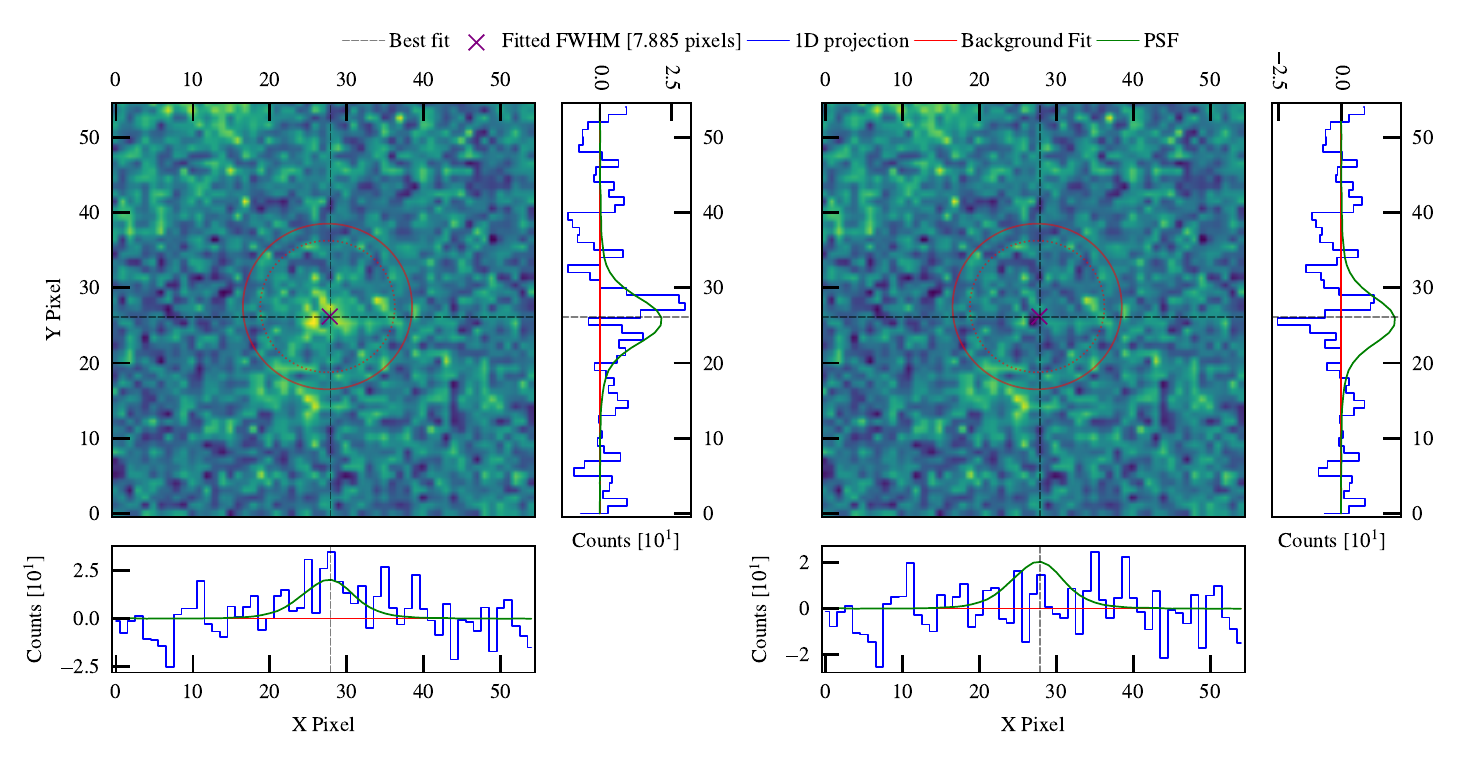}
\includegraphics[width=1\textwidth]{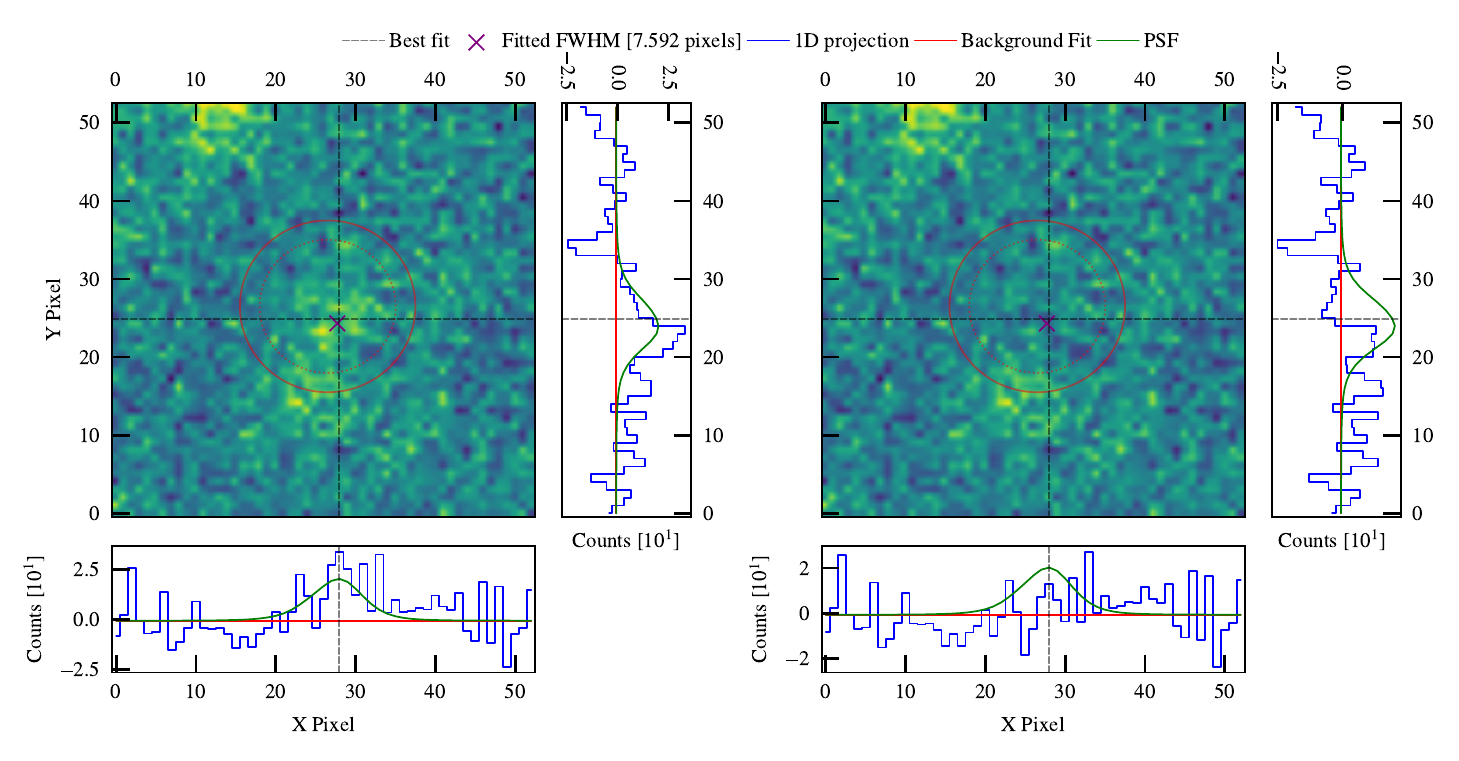}
\caption{Same as Figure \ref{fig:AT2020ddu_keck} but for AT~2020oor.}
\label{fig:AT2020oor_keck}
\end{figure*}

\restartappendixnumbering
We obtained imaging at the positions of AT~2019lbw, AT~2020ddu, and AT~2020oor with Keck~I/LRIS. The PSF subtractions at these positions generated by \auto~(see Section \ref{sec:2019lbw}, \ref{2022ddu}, and \ref{sec:2020oor}) are presented in Figures \ref{fig:AT2019lbw_keck}--\ref{fig:AT2020oor_keck}.

{}


\begin{thebibliography}{}
\bibitem[LIGO Scientific Collaboration (2017)]{2017ApJ...848L..12A} Abbott, B.~P., Abbott, R., Abbott, T.~D., et al.\ 2017, \apjl, 848, L12. doi:10.3847/2041-8213/aa91c9
\bibitem[Alam et al.(2015)]{2015ApJS..219...12A} Alam, S., Albareti, F.~D., Allende Prieto, C., et al.\ 2015, \apjs, 219, 12. doi:10.1088/0067-0049/219/1/12
\bibitem[Anand et al.(2021)]{2021NatAs...5...46A} Anand, S., Coughlin, M.~W., Kasliwal, M.~M., et al.\ 2021, Nature Astronomy, 5, 46. doi:10.1038/s41550-020-1183-3
\bibitem[Andreoni et al.(2017)]{2017PASA...34...69A} Andreoni, I., Ackley, K., Cooke, J., et al.\ 2017, \pasa, 34, e069. doi:10.1017/pasa.2017.65
\bibitem[Andreoni et al.(2020)]{2020ApJ...904..155A} Andreoni, I., Kool, E.~C., Sagu{\'e}s Carracedo, A., et al.\ 2020, \apj, 904, 155. doi:10.3847/1538-4357/abbf4c
\bibitem[Andreoni et al.(2021)]{2021ApJ...918...63A} Andreoni, I., Coughlin, M.~W., Kool, E.~C., et al.\ 2021, \apj, 918, 63. doi:10.3847/1538-4357/ac0bc7
\bibitem[Arcavi et al.(2016)]{2016ApJ...819...35A} Arcavi, I., Wolf, W.~M., Howell, D.~A., et al.\ 2016, \apj, 819, 35. doi:10.3847/0004-637X/819/1/35
\bibitem[Arcavi et al.(2017)]{2017Natur.551...64A} Arcavi, I., Hosseinzadeh, G., Howell, D.~A., et al.\ 2017, \nat, 551, 64. doi:10.1038/nature24291
\bibitem[Arcavi(2018)]{2018ApJ...855L..23A} Arcavi, I.\ 2018, \apjl, 855, L23. doi:10.3847/2041-8213/aab267
\bibitem[Astier et al.(2006)]{2006A&A...447...31A} Astier, P., Guy, J., Regnault, N., et al.\ 2006, \aap, 447, 31. doi:10.1051/0004-6361:20054185
\bibitem[Astropy Collaboration et al.(2013)]{2013A&A...558A..33A} Astropy Collaboration, Robitaille, T.~P., Tollerud, E.~J., et al.\ 2013, \aap, 558, A33. doi:10.1051/0004-6361/201322068
\bibitem[Barnes \& Kasen(2013)]{2013ApJ...775...18B} Barnes, J. \& Kasen, D.\ 2013, \apj, 775, 18. doi:10.1088/0004-637X/775/1/18
\bibitem[Bellm et al.(2019)]{2019PASP..131f8003B} Bellm, E.~C., Kulkarni, S.~R., Barlow, T., et al.\ 2019, \pasp, 131, 068003. doi:10.1088/1538-3873/ab0c2a
\bibitem[Bildsten et al.(2007)]{2007ApJ...662L..95B} Bildsten, L., Shen, K.~J., Weinberg, N.~N., et al.\ 2007, \apjl, 662, L95. doi:10.1086/519489
\bibitem[Bouy et al.(2013)]{2013A&A...554A.101B} Bouy, H., Bertin, E., Moraux, E., et al.\ 2013, \aap, 554, A101. doi:10.1051/0004-6361/201220748
\bibitem[Breedt et al.(2014)]{2014MNRAS.443.3174B} Breedt, E., G{\"a}nsicke, B.~T., Drake, A.~J., et al.\ 2014, \mnras, 443, 3174. doi:10.1093/mnras/stu1377
\bibitem[\protect\citeauthoryear{Brennan \& Fraser}{2022}]{2022A&A...667A..62B} Brennan S.~J., Fraser M., 2022, A\&A, 667, A62. doi:10.1051/0004-6361/202243067\bibitem[Chambers et al.(2016)]{2016arXiv161205560C} Chambers, K.~C., Magnier, E.~A., Metcalfe, N., et al.\ 2016, arXiv:1612.05560
\bibitem[Bulla(2019)]{2019MNRAS.489.5037B} Bulla, M.\ 2019, \mnras, 489, 5037. doi:10.1093/mnras/stz2495
\bibitem[Caswell et al.(2021)]{2021zndo....592536C} Caswell, T.~A., Droettboom, M., Lee, A., et al.\ 2021, Zenodo
\bibitem[Chornock et al.(2017)]{2017ApJ...848L..19C} Chornock, R., Berger, E., Kasen, D., et al.\ 2017, \apjl, 848, L19. doi:10.3847/2041-8213/aa905c
\bibitem[Coppejans et al.(2016)]{2016MNRAS.456.4441C} Coppejans, D.~L., K{\"o}rding, E.~G., Knigge, C., et al.\ 2016, \mnras, 456, 4441. doi:10.1093/mnras/stv2921
\bibitem[Coulter et al.(2017)]{2017Sci...358.1556C} Coulter, D.~A., Foley, R.~J., Kilpatrick, C.~D., et al.\ 2017, Science, 358, 1556. doi:10.1126/science.aap9811
\bibitem[Cowperthwaite et al.(2017)]{2017ApJ...848L..17C} Cowperthwaite, P.~S., Berger, E., Villar, V.~A., et al.\ 2017, \apjl, 848, L17. doi:10.3847/2041-8213/aa8fc7
\bibitem[Dahiwale \& Fremling(2019)]{2019TNSCR2407....1D} Dahiwale, A. \& Fremling, C.\ 2019, Transient Name Server Classification Report, 2019-2407
\bibitem[Dahiwale \& Fremling(2020)]{2020TNSCR2139....1D} Dahiwale, A. \& Fremling, C.\ 2020, Transient Name Server Classification Report, 2020-2139
\bibitem[D{\'a}lya et al.(2018)]{2018MNRAS.479.2374D} D{\'a}lya, G., Galg{\'o}czi, G., Dobos, L., et al.\ 2018, \mnras, 479, 2374. doi:10.1093/mnras/sty1703
\bibitem[Dey et al.(2019)]{2019AJ....157..168D} Dey, A., Schlegel, D.~J., Lang, D., et al.\ 2019, \aj, 157, 168. doi:10.3847/1538-3881/ab089d
\bibitem[D{\'\i}az et al.(2020)]{2020TNSCR3836....1D} D{\'\i}az, M., Morrell, N., Phillips, M., et al.\ 2020, Transient Name Server Classification Report, 2020-3836
\bibitem[D{\'\i}az et al.(2017)]{2017ApJ...848L..29D} D{\'\i}az, M.~C., Macri, L.~M., Garcia Lambas, D., et al.\ 2017, \apjl, 848, L29. doi:10.3847/2041-8213/aa9060
\bibitem[Dietrich et al.(2020)]{2020Sci...370.1450D} Dietrich, T., Coughlin, M.~W., Pang, P.~T.~H., et al.\ 2020, Science, 370, 1450. doi:10.1126/science.abb4317
\bibitem[Do(2021)]{2021TNSCR2033....1D} Do, A.\ 2021, Transient Name Server Classification Report, 2021-2033
\bibitem[Drake et al.(2009)]{2009ApJ...696..870D} Drake, A.~J., Djorgovski, S.~G., Mahabal, A., et al.\ 2009, \apj, 696, 870. doi:10.1088/0004-637X/696/1/870
\bibitem[Drake et al.(2014)]{2014MNRAS.441.1186D} Drake, A.~J., G{\"a}nsicke, B.~T., Djorgovski, S.~G., et al.\ 2014, \mnras, 441, 1186. doi:10.1093/mnras/stu639
\bibitem[Drout et al.(2011)]{2011ApJ...741...97D} Drout, M.~R., Soderberg, A.~M., Gal-Yam, A., et al.\ 2011, \apj, 741, 97. doi:10.1088/0004-637X/741/2/97
\bibitem[Drout et al.(2013)]{2013ApJ...774...58D} Drout, M.~R., Soderberg, A.~M., Mazzali, P.~A., et al.\ 2013, \apj, 774, 58. doi:10.1088/0004-637X/774/1/58
\bibitem[Drout et al.(2014)]{2014ApJ...794...23D} Drout, M.~R., Chornock, R., Soderberg, A.~M., et al.\ 2014, \apj, 794, 23. doi:10.1088/0004-637X/794/1/23
\bibitem[Drout et al.(2017)]{2017Sci...358.1570D} Drout, M.~R., Piro, A.~L., Shappee, B.~J., et al.\ 2017, Science, 358, 1570. doi:10.1126/science.aaq0049
\bibitem[Evans et al.(2017)]{2017Sci...358.1565E} Evans, P.~A., Cenko, S.~B., Kennea, J.~A., et al.\ 2017, Science, 358, 1565. doi:10.1126/science.aap9580
\bibitem[Feindt et al.(2019)]{2019JCAP...10..005F} Feindt, U., Nordin, J., Rigault, M., et al.\ 2019, \jcap, 2019, 005. doi:10.1088/1475-7516/2019/10/005
\bibitem[Flaugher(2005)]{2005IJMPA..20.3121F} Flaugher, B.\ 2005, International Journal of Modern Physics A, 20, 3121. doi:10.1142/S0217751X05025917
\bibitem[Forster et al.(2020)]{2020TNSTR2075....1F} Forster, F., Bauer, F.~E., Galbany, L., et al.\ 2020, Transient Name Server Discovery Report, 2020-2075
\bibitem[F{\"o}rster et al.(2021)]{2021AJ....161..242F} F{\"o}rster, F., Cabrera-Vives, G., Castillo-Navarrete, E., et al.\ 2021, \aj, 161, 242. doi:10.3847/1538-3881/abe9bc
\bibitem[Fremling et al.(2016)]{2016A&A...593A..68F} Fremling, C., Sollerman, J., Taddia, F., et al.\ 2016, \aap, 593, A68. doi:10.1051/0004-6361/201628275
\bibitem[Fremling et al.(2018a)]{2018TNSCR2146....1F} Fremling, C., Dugas, A., \& Sharma, Y.\ 2018, Transient Name Server Classification Report, 2018-2146
\bibitem[Fremling et al.(2018b)]{2018TNSCR1404....1F} Fremling, C., Sharma, Y., \& Dugas, A.\ 2018, Transient Name Server Classification Report, 2018-1404
\bibitem[Fremling et al.(2020a)]{2020ApJ...895...32F} Fremling, C., Miller, A.~A., Sharma, Y., et al.\ 2020, \apj, 895, 32. doi:10.3847/1538-4357/ab8943
\bibitem[Fremling(2020b)]{2020TNSTR.598....1F} Fremling, C.\ 2020, Transient Name Server Discovery Report, 2020-598
\bibitem[Gal-Yam et al.(2021)]{2021TNSAN..76....1G} Gal-Yam, A., Yaron, O., Pastorello, A., et al.\ 2021, Transient Name Server AstroNote, 76
\bibitem[Gaia Collaboration et al.(2022)]{2022arXiv220800211G} Gaia Collaboration, Vallenari, A., Brown, A.~G.~A., et al.\ 2022, arXiv:2208.00211
\bibitem[Gehrels(1986)]{1986ApJ...303..336G} Gehrels, N.\ 1986, \apj, 303, 336. doi:10.1086/164079
\bibitem[Graham et al.(2019)]{2019PASP..131g8001G} Graham, M.~J., Kulkarni, S.~R., Bellm, E.~C., et al.\ 2019, \pasp, 131, 078001. doi:10.1088/1538-3873/ab006c
\bibitem[Gromadzki et al.(2018)]{2018TNSCR1339....1G} Gromadzki, M., Wevers, T., Pignata, G., et al.\ 2018, Transient Name Server Classification Report, 2018-1339
\bibitem[Ho et al.(2021)]{2021arXiv210508811H} Ho, A.~Y.~Q., Perley, D.~A., Gal-Yam, A., et al.\ 2021, arXiv:2105.08811
\bibitem[Hodgkin et al.(2019)]{2019TNSTR1270....1H} Hodgkin, S.~T., Breedt, E., Delgado, A., et al.\ 2019, Transient Name Server Discovery Report, 2019-1270
\bibitem[Hosseinzadeh et al.(2017)]{2017ApJ...836..158H} Hosseinzadeh, G., Arcavi, I., Valenti, S., et al.\ 2017, \apj, 836, 158. doi:10.3847/1538-4357/836/2/158
\bibitem[Hu et al.(2017)]{2017SciBu..62.1433H} Hu, L., Wu, X., Andreoni, I., et al.\ 2017, Science Bulletin, 62, 1433. doi:10.1016/j.scib.2017.10.006
\bibitem[Ivezi{\'c} et al.(2019)]{2019ApJ...873..111I} Ivezi{\'c}, {\v{Z}}., Kahn, S.~M., Tyson, J.~A., et al.\ 2019, \apj, 873, 111. doi:10.3847/1538-4357/ab042c
\bibitem[Kasliwal et al.(2010)]{2010ApJ...723L..98K} Kasliwal, M.~M., Kulkarni, S.~R., Gal-Yam, A., et al.\ 2010, \apjl, 723, L98. doi:10.1088/2041-8205/723/1/L98
\bibitem[Kasliwal et al.(2017)]{2017Sci...358.1559K} Kasliwal, M.~M., Nakar, E., Singer, L.~P., et al.\ 2017, Science, 358, 1559. doi:10.1126/science.aap9455
\bibitem[Kilpatrick et al.(2017)]{2017Sci...358.1583K} Kilpatrick, C.~D., Foley, R.~J., Kasen, D., et al.\ 2017, Science, 358, 1583. doi:10.1126/science.aaq0073
\bibitem[Kleiser \& Kasen(2014)]{2014MNRAS.438..318K} Kleiser, I.~K.~W. \& Kasen, D.\ 2014, \mnras, 438, 318. doi:10.1093/mnras/stt2191
\bibitem[Kleiser et al.(2018)]{2018MNRAS.475.3152K} Kleiser, I.~K.~W., Kasen, D., \& Duffell, P.~C.\ 2018, \mnras, 475, 3152. doi:10.1093/mnras/stx3321
\bibitem[Kochanek et al.(2017)]{2017PASP..129j4502K} Kochanek, C.~S., Shappee, B.~J., Stanek, K.~Z., et al.\ 2017, \pasp, 129, 104502. doi:10.1088/1538-3873/aa80d9
\bibitem[Kulkarni(2005)]{2005astro.ph.10256K} Kulkarni, S.~R.\ 2005, astro-ph/0510256. doi:10.48550/arXiv.astro-ph/0510256
\bibitem[Kupfer et al.(2021)]{2021MNRAS.505.1254K} Kupfer, T., Prince, T.~A., van Roestel, J., et al.\ 2021, \mnras, 505, 1254. doi:10.1093/mnras/stab1344

\bibitem[\protect\citeauthoryear{Lam et al.}{2021}]{2021arXiv211102127L} Lam M.~C., Smith R.~J., Arcavi I., Steele I.~A., Veitch-Michaelis J., Wyrzykowski L., 2021, arXiv, arXiv:2111.02127
\bibitem[\protect\citeauthoryear{Lam}{2023}]{aspired_zenodo} Lam M.~C., 2023, zndo, doi:10.5281/zenodo.4127294

\bibitem[Law et al.(2009)]{2009PASP..121.1395L} Law, N.~M., Kulkarni, S.~R., Dekany, R.~G., et al.\ 2009, \pasp, 121, 1395. doi:10.1086/648598
\bibitem[Li \& Paczy{\'n}ski(1998)]{1998ApJ...507L..59L} Li, L.-X. \& Paczy{\'n}ski, B.\ 1998, \apjl, 507, L59. doi:10.1086/311680
\bibitem[Lipunov et al.(2017)]{2017ApJ...850L...1L} Lipunov, V.~M., Gorbovskoy, E., Kornilov, V.~G., et al.\ 2017, \apjl, 850, L1. doi:10.3847/2041-8213/aa92c0
\bibitem[Lunnan et al.(2017)]{2017ApJ...836...60L} Lunnan, R., Kasliwal, M.~M., Cao, Y., et al.\ 2017, \apj, 836, 60. doi:10.3847/1538-4357/836/1/60
\bibitem[Mahabal et al.(2019)]{2019PASP..131c8002M} Mahabal, A., Rebbapragada, U., Walters, R., et al.\ 2019, \pasp, 131, 038002. doi:10.1088/1538-3873/aaf3fa
\bibitem[Masci et al.(2019)]{2019PASP..131a8003M} Masci, F.~J., Laher, R.~R., Rusholme, B., et al.\ 2019, \pasp, 131, 018003. doi:10.1088/1538-3873/aae8ac
\bibitem[McBrien et al.(2019)]{2019ApJ...885L..23M} McBrien, O.~R., Smartt, S.~J., Chen, T.-W., et al.\ 2019, \apjl, 885, L23. doi:10.3847/2041-8213/ab4dae
\bibitem[McBrien et al.(2021)]{2021MNRAS.500.4213M} McBrien, O.~R., Smartt, S.~J., Huber, M.~E., et al.\ 2021, \mnras, 500, 4213. doi:10.1093/mnras/staa3361
\bibitem[Metzger et al.(2010)]{2010MNRAS.406.2650M} Metzger, B.~D., Mart{\'\i}nez-Pinedo, G., Darbha, S., et al.\ 2010, \mnras, 406, 2650. doi:10.1111/j.1365-2966.2010.16864.x
\bibitem[Metzger \& Berger(2012)]{2012ApJ...746...48M} Metzger, B.~D. \& Berger, E.\ 2012, \apj, 746, 48. doi:10.1088/0004-637X/746/1/48
\bibitem[Metzger(2019)]{2019LRR....23....1M} Metzger, B.~D.\ 2019, Living Reviews in Relativity, 23, 1. doi:10.1007/s41114-019-0024-0
\bibitem[M{\"o}ller et al.(2021)]{2021MNRAS.501.3272M} M{\"o}ller, A., Peloton, J., Ishida, E.~E.~O., et al.\ 2021, \mnras, 501, 3272. doi:10.1093/mnras/staa3602
\bibitem[Munari(2012)]{2012JAVSO..40..582M} Munari, U.\ 2012, \jaavso, 40, 582
\bibitem[Nakar(2020)]{2020PhR...886....1N} Nakar, E.\ 2020, \physrep, 886, 1. doi:10.1016/j.physrep.2020.08.008
\bibitem[Narayan et al.(2018)]{2018ApJS..236....9N} Narayan, G., Zaidi, T., Soraisam, M.~D., et al.\ 2018, \apjs, 236, 9. doi:10.3847/1538-4365/aab781
\bibitem[Necker et al.(2022)]{2022MNRAS.516.2455N} Necker, J., de Jaeger, T., Stein, R., et al.\ 2022, \mnras, 516, 2455. doi:10.1093/mnras/stac2261
\bibitem[Nordin et al.(2019a)]{2019TNSTR1202....1N} Nordin, J., Brinnel, V., Giomi, M., et al.\ 2019, Transient Name Server Discovery Report, 2019-1202
\bibitem[Nordin et al.(2019b)]{2019TNSTR2579....1N} Nordin, J., Brinnel, V., Giomi, M., et al.\ 2019, Transient Name Server Discovery Report, 2019-2579
\bibitem[Nordin et al.(2020a)]{2020TNSTR1797....1N} Nordin, J., Brinnel, V., Giomi, M., et al.\ 2020, Transient Name Server Discovery Report, 2020-1797
\bibitem[Nordin et al.(2020b)]{2020TNSTR2113....1N} Nordin, J., Brinnel, V., Giomi, M., et al.\ 2020, Transient Name Server Discovery Report, 2020-2113
\bibitem[Ofek et al.(2010)]{2010ApJ...724.1396O} Ofek, E.~O., Rabinak, I., Neill, J.~D., et al.\ 2010, \apj, 724, 1396. doi:10.1088/0004-637X/724/2/1396
\bibitem[Oke \& Gunn(1983)]{1983ApJ...266..713O} Oke, J.~B. \& Gunn, J.~E.\ 1983, \apj, 266, 713. doi:10.1086/160817
\bibitem[Oke et al.(1995)]{1995PASP..107..375O} Oke, J.~B., Cohen, J.~G., Carr, M., et al.\ 1995, \pasp, 107, 375. doi:10.1086/133562
\bibitem[Pastorello et al.(2021a)]{2021TNSCR.511....1P} Pastorello, A., Vogl, C., Taubenberger, S., et al.\ 2021, Transient Name Server Classification Report, 2021-511
\bibitem[Pastorello et al.(2021b)]{2021TNSAN..71....1P} Pastorello, A., Vogl, C., Taubenberger, S., et al.\ 2021, Transient Name Server AstroNote, 71
\bibitem[Payne(2021)]{2021TNSCR3242....1P} Payne, A.~V.\ 2021, Transient Name Server Classification Report, 2021-3242
\bibitem[Perley et al.(2019a)]{2019MNRAS.484.1031P} Perley, D.~A., Mazzali, P.~A., Yan, L., et al.\ 2019, \mnras, 484, 1031. doi:10.1093/mnras/sty3420
\bibitem[Perley (2019b)]{2019PASP..131h4503P} Perley, D.~A.\ 2019, \pasp, 131, 084503. doi:10.1088/1538-3873/ab215d
\bibitem[Perley et al.(2020)]{2020ApJ...904...35P} Perley, D.~A., Fremling, C., Sollerman, J., et al.\ 2020, \apj, 904, 35. doi:10.3847/1538-4357/abbd98
\bibitem[Pian et al.(2017)]{2017Natur.551...67P} Pian, E., D'Avanzo, P., Benetti, S., et al.\ 2017, \nat, 551, 67. doi:10.1038/nature24298
\bibitem[(2020)]{2020A&A...641A...6P} Planck Collaboration, Aghanim, N., Akrami, Y., et al.\ 2020, \aap, 641, A6. doi:10.1051/0004-6361/201833910
\bibitem[Poznanski et al.(2010)]{2010Sci...327...58P} Poznanski, D., Chornock, R., Nugent, P.~E., et al.\ 2010, Science, 327, 58. doi:10.1126/science.1181709
\bibitem[Poznanski et al.(2012)]{2012MNRAS.426.1465P} Poznanski, D., Prochaska, J.~X., \& Bloom, J.~S.\ 2012, \mnras, 426, 1465. doi:10.1111/j.1365-2966.2012.21796.x
\bibitem[Pursiainen et al.(2018)]{2018MNRAS.481..894P} Pursiainen, M., Childress, M., Smith, M., et al.\ 2018, \mnras, 481, 894. doi:10.1093/mnras/sty2309
\bibitem[Rastinejad et al.(2022)]{2022Natur.612..223R} Rastinejad, J.~C., Gompertz, B.~P., Levan, A.~J., et al.\ 2022, \nat, 612, 223. doi:10.1038/s41586-022-05390-w
\bibitem[Rau et al.(2009)]{2009PASP..121.1334R} Rau, A., Kulkarni, S.~R., Law, N.~M., et al.\ 2009, \pasp, 121, 1334. doi:10.1086/605911
\bibitem[Reback et al.(2022)]{2022zndo...3509134R} Reback, J., jbrockmendel, McKinney, W., et al.\ 2022, Zenodo
\bibitem[Rest et al.(2018)]{2018NatAs...2..307R} Rest, A., Garnavich, P.~M., Khatami, D., et al.\ 2018, Nature Astronomy, 2, 307. doi:10.1038/s41550-018-0423-2
\bibitem[Richmond et al.(1996)]{1996AJ....111..327R} Richmond, M.~W., van Dyk, S.~D., Ho, W., et al.\ 1996, \aj, 111, 327. doi:10.1086/117785
\bibitem[Rockosi et al.(2010)]{2010SPIE.7735E..0RR} Rockosi, C., Stover, R., Kibrick, R., et al.\ 2010, \procspie, 7735, 77350R. doi:10.1117/12.856818
\bibitem[Rosswog(2005)]{2005ApJ...634.1202R} Rosswog, S.\ 2005, \apj, 634, 1202. doi:10.1086/497062
\bibitem[Schlafly \& Finkbeiner(2011)]{2011ApJ...737..103S} Schlafly, E.~F., \& Finkbeiner, D.~P.\ 2011, \apj, 737, 103
\bibitem[Schlegel et al.(1998)]{1998ApJ...500..525S} Schlegel, D.~J., Finkbeiner, D.~P., \& Davis, M.\ 1998, \apj, 500, 525. doi:10.1086/305772
\bibitem[Shappee et al.(2014)]{2014ApJ...788...48S} Shappee, B.~J., Prieto, J.~L., Grupe, D., et al.\ 2014, \apj, 788, 48. doi:10.1088/0004-637X/788/1/48
\bibitem[Shappee et al.(2017)]{2017Sci...358.1574S} Shappee, B.~J., Simon, J.~D., Drout, M.~R., et al.\ 2017, Science, 358, 1574. doi:10.1126/science.aaq0186
\bibitem[Smartt et al.(2017)]{2017Natur.551...75S} Smartt, S.~J., Chen, T.-W., Jerkstrand, A., et al.\ 2017, \nat, 551, 75. doi:10.1038/nature24303
\bibitem[Smith et al.(2019)]{2019RNAAS...3...26S} Smith, K.~W., Williams, R.~D., Young, D.~R., et al.\ 2019, Research Notes of the American Astronomical Society, 3, 26. doi:10.3847/2515-5172/ab020f
\bibitem[Smith et al.(2020)]{2020PASP..132h5002S} Smith, K.~W., Smartt, S.~J., Young, D.~R., et al.\ 2020, \pasp, 132, 085002. doi:10.1088/1538-3873/ab936e
\bibitem[Strope et al.(2010)]{2010AJ....140...34S} Strope, R.~J., Schaefer, B.~E., \& Henden, A.~A.\ 2010, \aj, 140, 34. doi:10.1088/0004-6256/140/1/34
\bibitem[Tachibana \& Miller(2018)]{2018PASP..130l8001T} Tachibana, Y. \& Miller, A.~A.\ 2018, \pasp, 130, 128001. doi:10.1088/1538-3873/aae3d9
\bibitem[Tago et al.(2010)]{2010A&A...514A.102T} Tago, E., Saar, E., Tempel, E., et al.\ 2010, \aap, 514, A102. doi:10.1051/0004-6361/200913687
\bibitem[Tanaka \& Hotokezaka(2013)]{2013ApJ...775..113T} Tanaka, M. \& Hotokezaka, K.\ 2013, \apj, 775, 113. doi:10.1088/0004-637X/775/2/113
\bibitem[Tanvir et al.(2017)]{2017ApJ...848L..27T} Tanvir, N.~R., Levan, A.~J., Gonz{\'a}lez-Fern{\'a}ndez, C., et al.\ 2017, \apjl, 848, L27. doi:10.3847/2041-8213/aa90b6
\bibitem[Tauris et al.(2013)]{2013ApJ...778L..23T} Tauris, T.~M., Langer, N., Moriya, T.~J., et al.\ 2013, \apjl, 778, L23. doi:10.1088/2041-8205/778/2/L23
\bibitem[Tonry(2011)]{2011PASP..123...58T} Tonry, J.~L.\ 2011, \pasp, 123, 58. doi:10.1086/657997
\bibitem[Tonry et al.(2018)]{2018PASP..130f4505T} Tonry, J.~L., Denneau, L., Heinze, A.~N., et al.\ 2018, \pasp, 130, 064505. doi:10.1088/1538-3873/aabadf
\bibitem[Troja et al.(2022)]{2022Natur.612..228T} Troja, E., Fryer, C.~L., O'Connor, B., et al.\ 2022, \nat, 612, 228. doi:10.1038/s41586-022-05327-3
\bibitem[Utsumi et al.(2017)]{2017PASJ...69..101U} Utsumi, Y., Tanaka, M., Tominaga, N., et al.\ 2017, \pasj, 69, 101. doi:10.1093/pasj/psx118
\bibitem[Valenti et al.(2017)]{2017ApJ...848L..24V} Valenti, S., Sand, D.~J., Yang, S., et al.\ 2017, \apjl, 848, L24. doi:10.3847/2041-8213/aa8edf
\bibitem[Villar et al.(2017)]{2017ApJ...851L..21V} Villar, V.~A., Guillochon, J., Berger, E., et al.\ 2017, \apjl, 851, L21. doi:10.3847/2041-8213/aa9c84
\bibitem[Waxman et al.(2022)]{2022arXiv220610710W} Waxman, E., Ofek, E.~O., \& Kushnir, D.\ 2022, arXiv:2206.10710
\bibitem[Wevers et al.(2018)]{2018TNSCR1397....1W} Wevers, T., Gromadzki, M., Lyman, J., et al.\ 2018, Transient Name Server Classification Report, 2018-1397
\bibitem[Whitesides et al.(2017)]{2017ApJ...851..107W} Whitesides, L., Lunnan, R., Kasliwal, M.~M., et al.\ 2017, \apj, 851, 107. doi:10.3847/1538-4357/aa99de
\bibitem[Xiang et al.(2021)]{2021ApJ...910...42X} Xiang, D., Wang, X., Lin, W., et al.\ 2021, \apj, 910, 42. doi:10.3847/1538-4357/abdeba
\bibitem[York et al.(2000)]{2000AJ....120.1579Y} York, D.~G., Adelman, J., Anderson, J.~E., et al.\ 2000, \aj, 120, 1579. doi:10.1086/301513
\bibitem[Zou et al.(2019)]{2019ApJS..245....4Z} Zou, H., Zhou, X., Fan, X., et al.\ 2019, \apjs, 245, 4. doi:10.3847/1538-4365/ab48e8

\end{thebibliography}
\end{document}